\documentclass[%
 reprint,
nofootinbib,
 amsmath,amssymb,
 aps,
floatfix,
]{revtex4-2}

\usepackage{graphicx}
\usepackage{dcolumn}
\usepackage{bm}
\usepackage{caption}
\usepackage{aas_macros}
\usepackage{hyperref}
\usepackage{xcolor}
\usepackage{gensymb}
\usepackage{subcaption}

\begin{document}

\preprint{APS/123-QED}

\title{A Black-Hole Excision Scheme for General Relativistic Core-Collapse Supernova Simulations}

\author{Bailey Sykes}
 \email{bailey.sykes@monash.edu}
\author{Bernhard Mueller}%
 \email{bernhard.mueller@monash.edu}
\affiliation{%
 School of Physics and Astronomy, Monash University, Clayton, VIC, 3010, Australia
}%
\author{Isabel Cordero-Carri\'on}
\affiliation{Departamento de Matem\'aticas, Universitat de Val\`encia, 46100 Burjassot, Val\`encia, Spain}
\author{Pablo Cerd\'a-Dur\'an}
\affiliation{Departamento de Astronom\'ia y Astrof\'isica, Universitat de Val\`encia, 46100 Burjassot, Val\`encia, Spain
}%
\author{J\'er\^ome Novak}
\affiliation{Laboratoire Univers et Théories, Observatoire de Paris, Université PSL, CNRS, Université de Paris-Cité, 92190 Meudon, France}%
\date{\today}

\begin{abstract}
Fallback supernovae and the collapsar scenario for long-gamma ray burst and hypernovae have received considerable interest as pathways to black-hole formation and extreme transient events. Consistent simulations of these scenarios require a general relativistic treatment and need to deal appropriately with the formation of a singularity. Free evolution schemes for the Einstein equations can handle the formation of black holes by means of excision or puncture schemes. However, in constrained schemes, which offer distinct advantages in long-term numerical stability in stellar collapse simulations over well above $10^{4}$ light-crossing time scales, the dynamical treatment of black-hole spacetimes is more challenging. Building on previous work on excision in conformally flat spacetimes, we here present the implementation of a black-hole excision scheme for supernova simulations with the \textsc{CoCoNuT-FMT} neutrino transport code. We describe in detail a choice of boundary conditions that ensures long-time numerical stability, and also address upgrades to the hydrodynamics solver that are required to stably evolve the relativistic accretion flow onto the black hole. The scheme is currently limited to a spherically symmetric metric, but the hydrodynamics can be treated multi-dimensionally. For demonstration, we present a spherically symmetric simulation of black-hole formation in an $85 M_\odot$ star, as well as a two-dimensional simulation of the fallback explosion of the same progenitor. These extend past $9 \, \textnormal{s}$ and $0.3 \, \textnormal{s}$ after black-hole formation, respectively.

\end{abstract}

\maketitle


\section{\label{sec:intro} Introduction}

Typically the final iron core collapse of massive stars eventually results in the formation of a neutron star
and a supernova explosion~\citep{Baade_Zwicky:1934}. In a substantial fraction of massive stars, however, iron core collapse will lead to black hole formation. Observations of supernova progenitors indicate that red supergiants with initial masses above $15\texttt{-}18 M_\odot$ will collapse quietly to black holes \citep{Smartt:2009,Smartt:2015}, and there is now
even direct evidence for the disappearance of a red supergiant of $\mathord{\sim} 25M_\odot$ \citep{Adams_Kochanek_Gerke_Stanek_Dai:2017}.
In some cases, black hole formation may not proceed quietly, however. It has long been theorized that
the formation of black holes with an accretion disk during the collapse of rapidly rotating massive stars may lead to powerful ``hypernova'' explosions and long gamma-ray bursts \citep{MacFadyen_Woosley:1999,Woosley_Bloom:2006} (collapsar scenario). Furthermore,
black holes may be formed, regardless of progenitor rotation, if an incipient supernova explosion is not
sufficiently energetic to eject the entire stellar envelope and a substantial fraction of the envelope undergoes fallback onto the neutron star during some stage of the explosion. Recently, a number of multi-dimensional 
supernova simulations have followed the long-term evolution of neutrino-driven explosions in massive progenitors
to study this fallback scenario \citep{Chan_Mueller_Heger_Pakmor_Springel:2018,Chan_Mueller:2020,stockinger_20,Rahman_Janka_Stockinger_Woosley:2021}.

Simulations of black-hole forming supernovae are technically challenging, however. In addition to the usual numerical challenges -- multi-dimensional fluid flow including magnetic fields and neutrino transport -- a general relativistic treatment of the spacetime is required to consistently follow the evolution of the supernova core up to, through, and beyond black hole formation. Furthermore, 
while a judicious choice of slicing conditions may
be sufficient to follow the accretion onto the black hole for some time after its formation \citep{Rahman_Janka_Stockinger_Woosley:2021},
special techniques such as excision \citep{Thornburg:1993} or puncture methods \citep{Brandt_Brugmann:1997} are generally required to
avoid problems with the central singularity and follow the accretion onto the black hole on longer time scales. As a simpler alternative, one can treat the long-term evolution after black hole formation in Newtonian gravity \citep{Chan_Mueller_Heger_Pakmor_Springel:2018,Chan_Mueller:2020}, perhaps in combination with a pseudo-relativistic potential \citep{Artemova_Bjoernsson_Novikov:1996, Marek_Dimmelmeier_Janka_Muller_Buras:2006},
but this approximate approach is not well suited
during the dynamical collapse phase and may not be accurate enough for treating collapsar disks and, in particular, fast relativistic outflows. A rigorous relativistic treatment is also required for accurate predictions of the gravitational wave signal from black-hole formation in core-collapse supernovae.

For this reason, there have only been few efforts
to consistently simulate collapsars and fallback supernovae through black hole formation in full general relativity. Many simulations of fallback supernovae stop when a black hole is formed, regardless of whether the spacetime was treated in full general relativity 
\citep{Kuroda_et_al:2018} or using approximate pseudo-Newtonian gravity \citep{Pan_Liebendorfer_Couch_Thielemann:2018, Pan_Liebendorfer_Couch_Thielemann:2021} up to that point. \citet{Chan_Mueller_Heger_Pakmor_Springel:2018,Chan_Mueller:2020} mapped from a relativistic simulation (in the conformally flat approximation \citep{Isenberg:2008,Wilson_Mathews_Marronetti:1996,CorderoCarrion_et_al:2009}) up to black hole formation to a Newtonian moving mesh code to follow the long-term evolution of fallback supernovae. Two studies have presented axisymmetric (2D) general relativistic simulations beyond black hole formation based on a puncture method, employing
either multi-group flux-limited diffusion or
\citep{Rahman_Janka_Stockinger_Woosley:2021} or
a simpler, gray transport scheme \citep{Fujibayashi_et_al:2021}. 
While models of collapsars and long gamma-ray bursts
\citep[e.g.][]{Komissarov_Barkov:2007, Barkov_Komissarov:2008, Siegel_Barnes_Metzger:2019}.
have long moved beyond the use of modified Newtonian gravity in early studies \citep{MacFadyen_Woosley:1999}, they commonly bypass the pre-collapse phase altogether and insert a black hole at the beginning.

In order to circumvent the problem of numerical
singularities or grid stretching in singularity avoiding slicings, two classes of methods have
been developed to evolve black hole spacetimes in numerical relativity. Puncture methods \citep{Pretorius:2007,Brandt_Brugmann:1997} factor out the singular parts of the spacetime from the regular parts. The singular parts are dealt with analytically while the regular parts are allowed to evolve numerically \citep{Centrella_Baker_Kelly_vanMeter:2010,Alcubierre:2008}. On the other hand, excision techniques involve removing a section of the spacetime and imposing boundary conditions on the excised surface \citep{Thornburg:1993, Baumgarte_Shapiro:2010}. 
Applications of puncture and excision schemes have largely been confined to hyperbolic evolution schemes.

Since core-collapse supernovae typically need to cover several $10^{4}$ light-crossing time scales of the compact object or more, it
has been popular to resort to 
the conformal flatness condition as an elliptic formulation of the Einstein equations for this problem \citep{Dimmelmeier_Font_Mueller:2002, Mueller_Janka_Dimmelmeier:2010}. While CFC is exact only in spherical symmetry, it enables stable long-term evolution by solving the elliptic constraint equations directly. The CFC approximation is also a stepping stone towards the fully constrained formalism (FCF) \citep{Bonazzola_Gourgoulhon_Grandclement_Novak:2004}, a non-approximate elliptic-hyperbolic formulation of the Einstein equations in the generalized Dirac gauge. Because of their stability properties and suitability for long-time simulations, 
adapting such constrained schemes to black hole spacetimes is a major desideratum. However, the elliptic nature of the constraint equations poses challenges in 
formulating excision or puncture schemes for constrained formulations. Building on earlier work around excision for the black hole initial data problem~\citep{Jaramillo_Gourgoulhon_Marugan:2004, Jaramillo:2008, Vasset_Novak_Jaramillo:2009}, an excision scheme based on CFC/FCF was formulated for spherically symmetric spacetimes by
\citet{CorderoCarrion_Vasset_Novak_Jaramillo:2014} and applied to several test cases (static black holes, scalar field collapse, collapse of an isolated neutron star). However, the scheme is yet to be applied in full core-collapse supernova simulations with neutrino transport.

In this paper, we implement a modified version of the excision scheme of \citet{CorderoCarrion_Vasset_Novak_Jaramillo:2014} in the \textsc{CoCoNuT-FMT} supernova code.
We also present results from simulations of black-hole formation during the collapse of a massive star in spherical symmetry and in axisymmetry (with a spherically symmetric metric) to demonstrate the excision scheme in practice.

Our paper is structure as follows.
In Section~\ref{sec:GR}, we review the CFC approximation and describe the boundary conditions for the excision scheme, including modifications from the original formulation of \citet{CorderoCarrion_Vasset_Novak_Jaramillo:2014}. Section~\ref{sec:metric_solver} provides further details on the numerical implementation of the boundary condition and the solution of the elliptic equations that are specific to \textsc{CoCoNuT-FMT} and more peripheral to the excision method per se. Section~\ref{sec:hydro} describes modifications to the hydrodynamics modules of the \textsc{CoCoNuT-FMT} code, which are required to stably model accretion flow across the black hole horizon. In Section~\ref{sec:tests}, we present the results of a spherically symmetric (1D) core-collapse supernova simulation of an  $85 M_{\odot}$ star. Finally, we show results from an axisymmetric (2D) simulation of a fallback supernovae (using the same progenitor) with the excision scheme
in Section \ref{sec:2d}.

Throughout this paper we use geometrized units: $G = c = 1$. Greek indices run from $0$ to $3$ while Latin indices run from $1$ to $3$.

\section{\label{sec:GR}Metric equations and Excision Scheme}

\subsection{\label{subsec:3+1}3+1 Formalism in the conformal flatness approximation}

For the numerical solution of dynamical problems, the Einstein equations must be formulated as an evolution problem using spacelike or null foliations. We start from the 3+1 formalism \citep{ADM:1962, Lichnerowicz:1944}, in which the four-dimensional spacetime is foliated into a continuous sequence of three-dimensional spacelike hypersurfaces. This foliation induces a metric with line element,
\begin{equation}
    ds^{2} = -\alpha^{2} dt^{2} + \gamma_{ij}(dx^{i} + \beta^{i}dt)(dx^{j} + \beta^{j} dt),
\end{equation}
where $\alpha$ is the lapse function, $\beta^{i}$ is the shift vector, and $\gamma_{ij}$ is the three-metric on the spacelike hypersurfaces. In this work, we adopt the approximation that the spatial metric is conformally flat (conformal flatness condition, CFC) \citep{Wilson_Mathews_Marronetti:1996} and spherically symmetric (although when assuming spherical symmetry, the former assumption ceases to be an approximation and is just a choice of isotropic spatial coordinates).\footnote{The CFC approximation is intimately
related to an exact mixed hyperbolic/elliptic formulation of the Einstein
equations in the generalized Dirac gauge known as the fully constrained formalism (FCF) \citep{Bonazzola_Gourgoulhon_Grandclement_Novak:2004}.}
The CFC approximation for relativistic gravity has been used extensively in various general relativistic hydrodynamics codes for many years \citep{Cook_Shapiro_Teukolsky:1996, Dimmelmeier_Font_Mueller:2002, Saijo:2005, Bauswein_Pulpillo_Janka_Goriely:2014}.

The CFC approximation
reduces the Einstein equations to a set of elliptic constraint equations.
In spherical polar coordinates, the CFC metric is given by,
\begin{equation}
    g_{\mu \nu} =
    \begin{pmatrix}
        -\alpha^{2} + \beta_{i} \beta^{i} & \beta_{r} & \beta_{\theta} & \beta_{\varphi} \\
        \beta_{1} & \phi^{4} & 0 & 0 \\
        \beta_{2} & 0 & \phi^{4} r^{2} & 0 \\
        \beta_{3} & 0 & 0 & \phi^{4} r^{2} \sin^{2}\theta
    \end{pmatrix}.
\end{equation}
To determine the conformal factor, $\phi$, in addition to the lapse, $\alpha$, and shift vector, $\beta^{i}$, the Hamiltonian constraint and momentum constraint are supplemented by a gauge condition, namely
maximal slicing, which requires that the trace of the extrinsic curvature $K_{ij}$ should vanish,
\begin{equation}
    K = \gamma^{ij} K_{ij} = 0.
\end{equation}

This results in a set of three coupled elliptic equations that describe the spacetime in the presence of a given stress-energy tensor, determined by the matter energy, momentum and stresses at a given time:
\begin{equation}
\label{eqn:elliptic_phi}
    \Delta \phi = -2 \pi \phi^{-1} \bigg[E^{*} + \frac{\phi^{6} K_{ij} K^{ij}}{16 \pi} \bigg],
\end{equation}
\begin{equation}
\label{eqn:elliptic_alpha}
    \Delta(\alpha \phi) = 2 \pi \alpha \phi^{-1} \bigg[ E^{*} + 2 S^{*} + \frac{7 \phi^{6} K_{ij} K^{ij}}{16 \pi} \bigg],
\end{equation}
\begin{equation}
\label{eqn:elliptic_beta}
    \Delta \beta^{i} + \frac{1}{3} \nabla^{i} \nabla_{j} \beta^{j} = 16 \pi \alpha \phi^{-2} (S^{*})^{i} + 2 \phi^{10} K^{ij} \nabla_{j} \frac{\alpha}{\phi^{6}},
\end{equation}
where $E^{*}$, $S^{*}$ and $(S^{*})^{i}$ are the conformally rescaled energy density, trace of the stress tensor, and momentum density respectively (which are conserved hydrodynamic variables). The flat-space Laplacian is denoted by $\Delta = f^{ij} \nabla_{i} \nabla_{j}$ where $f^{ij}$ is the flat space three-metric.
At spatial infinity, the metric is flat, meaning ${\phi \rightarrow 1}$, ${\alpha \rightarrow 1}$ and ${\beta^{i} \rightarrow 0}$. Also note that spherical symmetry causes the angular components of the shift to be zero, i.e., $\beta_{\theta} = \beta_{\varphi} = 0$, as well as the off-diagonals of the extrinsic curvature.
In our code, we use a fixed-point iteration \citep{Dimmelmeier_et_al:2005} based on a multipole expansion \citep{Muller_Steinmetz:1995} for the scalar and vector Poisson equations. The vector Poisson equation for the shift is reduced to scalar Poisson equations following \citet{Grandclement_Bonazzola_Gourgoulhon_Marck:2001}.
In general, the source terms on the right hand sides are dominated by terms containing hydrodynamic quantities; only for strong gravitational fields do the strongly non-linear curvature terms (e.g. $K_{ij}K^{ij}$) become significant. 

Prior to black hole formation, the extended CFC (XCFC) formalism of \citep{CorderoCarrion_et_al:2009} is used as it is robust against the uniqueness issue which sometimes plagues the original CFC approach (Equations \eqref{eqn:elliptic_phi}, \eqref{eqn:elliptic_alpha} and \eqref{eqn:elliptic_beta}). We prefer not to use the XCFC after black hole formation as it introduces an auxiliary vector $X^{i}$, which is difficult to define a boundary condition for. We observe no issues with the uniqueness of the solution in the simulation results.

\subsection{\label{subsec:BCs}Boundary conditions for the metric variables}

To extend our supernova code, \textsc{CoCoNuT-FMT}, to simulations beyond black hole formation, we follow \citet{CorderoCarrion_Vasset_Novak_Jaramillo:2014} and perform excision of a spherical region inside the PNS core by imposing suitable boundary conditions on the metric variables at the excision surface. This excision surface lies strictly within the apparent horizon once it forms. The boundary conditions for $\phi$, $\alpha$ and $\beta^{i}$ respectively are, in the original formulation,

\begin{equation}
    \partial_{t} \phi = \beta^{k} \nabla_{k} \phi + \frac{\phi}{6} \nabla_{k} \beta^{k}.
    \label{eqn:phi_bc}
\end{equation}
\begin{equation}
    \label{eqn:old_alpha_bc}
    \alpha = \frac{\phi^{6} (L\beta)^{ij} s_{i} s_{j}}{2 \hat{A}^{ij} s_{i} s_{j}},
\end{equation}
\begin{equation}
    \label{eqn:old_beta_bc}
    \beta^{i} s_{i} = \textnormal{constant},
\end{equation}

where $s_{i}$ is an  outward-directed spacelike unit vector normal to the excision surface, ${(L\beta)^{ij} := \mathcal{D}^{i} \beta^{j} + \mathcal{D}^{j} \beta^{i} - \frac{2}{3} f^{ij} \mathcal{D}_{k} \beta^{k}}$ and $\hat{A}^{ij}$ is the conformally rescaled extrinsic curvature, i.e., 
$\hat{A}^{ij}=\phi^{10} K^{ij}$. Unless otherwise noted, boundary conditions are implied to apply on the excision surface only. We evolve $\hat{A}^{ij}$ according to the time evolution equation as given by \citet{CorderoCarrion_CerdaDuran_MariaIbanez:2012}. Under the assumption of spherical symmetry only the component $\hat{A}^{rr}$ has an effect on the boundary conditions and it suffices to integrate,

\begin{equation}
\label{eqn:aij_evol}
\begin{split}
     \frac{\partial \hat{A}^{rr}}{\partial t} & = \beta^{r} \partial_{r} \hat{A}^{rr} + \frac{5}{3} \hat{A}^{rr} \bigg(\frac{2}{r} \beta^{r} + \partial_{r} \beta^{r} \bigg) - 2 \hat{A}^{rr} \partial_{r} \beta^{r} \\
    & + 2N \phi^{-6} (\hat{A}^{rr})^{2} - 8 \pi N \phi^{6} \bigg( \phi^{4} S^{rr} - \frac{S}{3} \bigg)  \\
    & + \frac{16}{3} N (\partial_{r} \phi)^{2} + \frac{16}{3} \phi  \partial_{r} \phi \partial_{r} N  \\
    & - \frac{2}{3} \big( \phi \Delta (N \phi) + N \phi \Delta \phi + 2 \partial_{r} \phi (\phi \partial_{r} N + N \partial_{r}\phi) \big),
\end{split}
\end{equation}

where $S^{rr}$ is the $rr$ component of the stress energy tensor, and $S$ is its trace. 

The set of equations \eqref{eqn:phi_bc} - \eqref{eqn:old_beta_bc} represent one possible gauge choice for the elliptic system. Equation \eqref{eqn:phi_bc} comes from the kinematic relations of \citet{Bonazzola_Gourgoulhon_Grandclement_Novak:2004}. Equation \eqref{eqn:old_alpha_bc} is the result of imposing conformal flatness, hence this relation (or some equivalent equation) is a required feature in the boundary conditions; see Equation (2.14) in \citet{CorderoCarrion_Vasset_Novak_Jaramillo:2014}. Equation \eqref{eqn:old_beta_bc} is a choice of gauge. It is possible to make an alternative gauge choice and, indeed, we find a modified gauge condition offers better numerical stability when combined with our hydrodynamics code. Specifically, we find the original boundary conditions to be sensitive to errors in the evolution of $\hat{A}^{ij}$, and that they can become unstable if $\hat{A}^{ij} \rightarrow 0$. Furthermore, while these boundary conditions do preserve the inward direction of the characteristics of the metric equations inside the apparent horizon, we find that null-characteristics are often produced outside the excision boundary, prompting the boundary to move outward over a duration of hundreds of milliseconds until it encompasses most of the grid (i.e. the apparent horizon becomes thousands of kilometers in radius). This behaviour is not desirable from a numerical point of view. Furthermore, we do not find good long-term conservation of the ADM mass with the original scheme if there is ongoing accretion onto the black hole over time scales of hundreds of milliseconds (as opposed to the test cases in \citep{CorderoCarrion_Vasset_Novak_Jaramillo:2014}, which quickly settled to a vacuum black hole space time).

In theory, the excision radius could be kept constant, which may produce a more realistic metric evolution for the original boundary conditions, however this would necessitate being able to handle the extremely relativistic flows near the centre of the grid. CoCoNuT is unable to accurately model hydrodynamics under these conditions and hence, the excision surface must be moved outward where possible, for any set of boundary conditions.  

To resolve these issues, we use new boundary conditions for the lapse and radial component of the shift vector on the excision surface. Respectively, these are,
\begin{equation}
    \frac{\alpha}{\phi^{2} \beta^{r}} = \textnormal{const.}
    \label{eqn:alpha_bc}
\end{equation}
\begin{equation}
    \tilde{\gamma}^{ik} \nabla_{k} \beta^{j} + \tilde{\gamma}^{kj} \nabla_{k} \beta^{i} - \frac{2}{3} \tilde{\gamma}^{ij} \nabla_{k} \beta^{k} = 2 \alpha \phi^{-6} \hat{A}^{ij}.
    \label{eqn:beta_bc}
\end{equation}
We also change the boundary condition for the conformal factor to explicitly guarantee conservation of the ADM mass, or rather an appropriate change of the ADM mass according to the energy flux through the \emph{outer} boundary of the grid. The boundary condition for $\phi$ thus changes from a boundary condition on the excision surface to a condition on the outer boundary,
\begin{equation}
    \label{eqn:phi_bc_adm}
    \phi'(R) = - \frac{M_\mathrm{ADM}}{2R^{2}},
\end{equation}
where $M_\mathrm{ADM}$ is the ADM mass (see \ref{subsec:adm}), $\phi'$ is the radial derivative of $\phi$ and $R = 10^{10} \, \textnormal{cm}$ is the radius of the outer boundary of the grid. An inner boundary is not required as the solution for $\phi$ is fully determined by the outer boundary condition and the sources on the grid. Equation \eqref{eqn:aij_evol} is still used to evolve $\hat{A}^{ij}$ in each timestep. The motivation behind each of these changes is explained in more detail in the following two sections.

\subsubsection{\label{subsubsec:lapse_shift_bc}Reformulated boundary condition: Lapse and shift}
The constraint from the hyperbolic sector imposed in Equation \eqref{eqn:old_alpha_bc} can be trivially rearranged to produce Equation \eqref{eqn:beta_bc}, which acts as a boundary condition for $\beta^{r}$. Some form of this condition must be present in any form of the boundary conditions. This leaves the freedom to manually set the lapse function, instead of the radial component of the shift vector. It is possible, and perhaps is most simple, to impose a constant value of the lapse on the excision boundary, $\alpha(R_\mathrm{AH}) = \textnormal{const.}$; set such that $\alpha$  is smooth and continuous in time over the transition from non-excised to excised regimes. While this is the obvious choice, we find that this gauge can fail to preserve the inward-pointing nature of outward-directed light rays inside the apparent horizon. This means that the apparent horizon can vanish if the metric functions, and specifically the boundary conditions, adjust such that,
\begin{equation}
    \frac{\alpha}{\phi^{2}} - \beta^{r} > 0
\end{equation}
inside the excised region. Here the LHS is just the coordinate velocity of radial outward-directed light rays. It is helpful then to consider an inner boundary condition for the lapse function where the ``outgoing'' null characteristic still point inward at the excision surface. We find that this can be achieved by maintaining the ratio $\alpha / (\phi^{2} \beta^{r})$, as per Equation \eqref{eqn:alpha_bc}, where its value is set, once again, by continuity over the transition to the excised metric. 

Tests with this boundary condition for the lapse function demonstrated that it prevents instabilities in the metric equations which can otherwise occur if the light cone at the excision surface is not properly pointing inward any more due to the adjustment of the metric as material accretes onto the black hole. It is also simple to implement, and does not involve any derivatives, which makes it efficient to compute and less prone to numerical accuracy problems under the restriction of finite spatial resolution. 

The shift vector is now constrained by a Robin boundary condition on the excision surface. We find no particular issues with imposing this boundary condition.

\subsubsection{\label{subsubsec:conformal_bc}Reformulated boundary condition: Conformal factor}

We now turn our attention to the boundary condition for the conformal factor, $\phi$. In the original scheme, $\phi$ on the excision boundary is governed by the time evolution equation \eqref{eqn:phi_bc}. This approach produces a stable simulation which performs well in almost every regard. However we found that, using Equation~\eqref{eqn:phi_bc}, the ADM mass may not be conserved well numerically. The ADM mass, as given by \citet{Gourgoulhon:2007}, is,
\begin{equation}
    \label{eqn:full_adm}
     M_{\textnormal{ADM}} = \frac{1}{16 \pi} \lim_{\mathcal{S} \rightarrow \infty} \oint_{\mathcal{S}} (\nabla^{j} \gamma_{ij} - \nabla_{i} (f^{kl} \gamma_{kl} )) s^{i} \sqrt{q} \,dy^{2} ,
\end{equation}
where $\mathcal{S}$ is typically taken as a sphere with induced surface element $\sqrt{q}\,d^{2}y$ and $s^{i}$ is an exterior-pointing unit normal to $S$.

Using Equation~\eqref{eqn:phi_bc} for the time evolution of $\phi$ on the excision surface produces a steady decline in the ADM mass post-excision, likely partially due to numerical inaccuracies in the metric or matter evolution. This is undesirable as it is analogous to a spurious loss of inertial mass, and hence may affect the dynamics of the collapsing star outside the excision region. This may impact, e.g., the trajectory of the shock in a prompt fallback scenario and generally invalidate the simulation results. 

As shown in more detail in Appendix~\ref{app:dADM_dt},
ADM mass conservation requires
a close cancellation of terms that
are of order $\beta^r M_\mathrm{ADM} /r_\mathrm{AH}^2 $ individually. With $\beta^r \sim 0.1$ at the horizon, even a small relative error of $10^{-3}\texttt{-}10^{-4}$ will lead to a change of $\mathcal{O}(M)$ over $10^{4}\texttt{-}10^{5}$ light-crossing time scales as required for realistic simulations of black-hole forming supernovae.
As a result, Equation~\eqref{eqn:phi_bc} is very sensitive to the discretisation error of numerical derivatives and hydrodynamic sources on the excision boundary. Specifically, ADM mass conservation depends critically on the numerical accuracy of the solution for $\beta^r$ and hence on the time evolution of $\hat{A}^{ij}$, which enters the boundary condition for $\beta^r$. Currently, a mix of upwind and centred discretisations are used for the evolution of $\hat{A}^{ij}$, with, for example, the first term treated with an upwind scheme. Tests with potentially more accurate discretisation schemes, such as
semi-implicit time integration or a higher-order, five-point stencil for the spatial derivatives in
Equation~\eqref{eqn:aij_evol}, did not yield any noticeable improvements, however.

The consequence of this is that we are not able to achieve the required accuracy in each of the three terms listed previously, and hence the evolution of $\phi$ is inaccurate. This causes the ADM mass to drift.

To prevent this, we instead adopt an ADM-mass conserving scheme for $\phi$, which explicitly holds the ADM mass constant (except for changes due to any energy flux through the outer boundary) by imposing a new boundary condition at the outer boundary of the grid. To obtain this boundary condition we start by rewriting Equation \eqref{eqn:full_adm} in a simpler form for the current case of a spherically symmetric CFC metric. In spherical symmetry, and with $\gamma_{ij}$ as previously identified, the expression is simplified to
\begin{equation}
    \label{eqn:lim_adm}
    M_{\rm ADM} = -2 \lim_{r \rightarrow \infty} \phi^{5} (\partial_{r} \phi) r^{2} .
\end{equation}
In practice, we cannot evaluate this beyond the largest radius in the grid, although in many cases the outer boundary serves as an acceptable proxy for a surface at infinity. A more rigorous solution is obtained by considering that, outside the grid, the energy density and curvature terms in Equation \eqref{eqn:elliptic_phi} are small and decay sufficiently fast such that an analytic solution is available: $\phi = c_{1}r^{-1} + c_{2}$. Matching this solution to the outer boundary of the grid with $c_{1}$ and setting $c_{2} = 1$ for the correct asymptotic behaviour enables us to approximate the ADM mass at spatial infinity:
\begin{equation}
    \label{eqn:extrap_ADM}
    M_\mathrm{ADM} = -2R^{2}\phi'(R),
\end{equation}
where $R$ is the radius of the outer boundary and $\phi'$ is the radial derivative of $\phi$.

Equation~\eqref{eqn:phi_bc_adm} is the new boundary condition, and acts to fix the radial derivative of $\phi$ on the outer boundary of the grid. This replaces the original boundary condition on the excision surface, Equation~\eqref{eqn:phi_bc}. There is no tension between this approach and Equation~\eqref{eqn:phi_bc} since the evolution of $\phi$ between the two methods must agree \emph{analytically}, but explicitly imposing ADM mass conservation helps to improve the numerical accuracy of the metric evolution due to better agreement with physical conservation laws.

The ADM mass of the star is expected to gradually decrease by a small amount due to energy losses from neutrinos streaming out of the star (and out of the numerical grid). Since the ADM mass can be expressed as a surface integral at spatial infinity, these neutrinos should always be included as they are always within this surface; however, since they cannot be accounted for once they leave the grid, and only the ADM mass on the grid is relevant, the ADM mass contribution of these neutrinos is removed once they cross the outer boundary. Similarly, material from outer layers of the star falling inwards through the outer boundary will contribute a small increase in ADM mass. In good approximation, this amounts to
\begin{equation}
    \label{eqn:delta_adm}
    \frac{\partial M_{\mathrm{ADM}}}{\partial t} = \oint_{\mathcal{S}} (\mathbf{F}_{\mathrm{n}} + \mathbf{F}_{\mathrm{m}})\cdot d\mathbf{A},
\end{equation}
where $\mathcal{S}$ is the outer surface of the numerical grid with surface element $dA$, neutrino flux $\mathbf{F}_{\mathrm{n}}$ and matter flux $\mathbf{F}_{\mathrm{m}}$. This neglects, for example, the contribution of mass outside the grid and small relativistic effects, but is justifiable because of the large radius of the grid boundary which allows us to apply familiar notions of energy conservation.
In practice, the neutrino flux across the outer grid boundary is small after black hole formation as the density of the infalling material outside the horizon (and hence the neutrino emission rate) is considerably lower than during the proto-neutron star phase. Nonetheless, the effect is included by decreasing the fixed ADM mass by the neutrino energy flux through the outer boundary. While the matter flux increases with time as the collapse of the outer layers accelerates, it is initially an order of magnitude less than the neutrino flux.

\subsection{\label{subsec:ah_finder}Apparent horizon finder}

The boundary conditions for the lapse and shift are enforced on a spherically symmetric shell -- the excision surface. The radius of this shell is determined by the requirement that it be inside the apparent horizon of the nascent black hole. The mathematical condition for  an apparent horizon  (marginally trapped surface) is that the expansion, $\Theta$, of outgoing null congruences should vanish,
\citep{Baumgarte_Shapiro:2003}, 
\begin{equation}
    \Theta(r_{\mathrm{AH}}) = \nabla_{i}s^{i} - K + s^{i} s^{j} K_{ij} = 0
\end{equation}
where $s^{i}$ is an outward pointing unit normal. This may be written  explicitly as an equation for $\alpha$, $\phi$, and $\beta^r$, recalling that maximal slicing is used so $K = 0$,
\begin{equation}
\label{eqn:AH_condition}
    \frac{4}{\phi^{3}} \frac{\partial \phi}{\partial r} + \frac{2}{\phi^{2} r} + \frac{2}{3 \alpha} \bigg( \frac{\partial \beta^{r}}{\partial r} - \frac{\beta^{r}}{r} \bigg) = 0,
\end{equation}
with $\Theta(r) < 0$ inside the apparent horizon surface. The excision surface is placed strictly inside the apparent horizon. Since the characteristics of the metric and fluid equations point inward, there are no consequences for not properly evolving the hydrodynamics inside the excision region. Indeed, other simulation codes may remove this region from the computational domain entirely. The requirements on boundary conditions for hydrodynamic variables are also less demanding as they primarily need only be numerically stable, with physical correctness being less important since the region outside the black hole is physically unaffected by boundary conditions for hyperoblic equations at the apparent horizon. These boundary conditions are discussed further in Section \ref{subsec:hydro_boundary}. 

The apparent horizon may move throughout the simulation, resulting in the excision surface travelling outward if the negative expansion condition is satisfied by a zone with greater radius. Enabling the excision surface to move can prevent numerical issues which arise when trying to simulate hydrodynamic flows too far inside the black hole. When moving the excision surface, the boundary conditions are redefined based on the new values of the metric variables on the excision surface, i.e., the conformal factor uses $\phi$ at the new radius as its initial value, the ratio $\alpha/(\phi^2 \beta^r)$ is reset using the metric quantities at the new radius, and the radial component of the shift is calculated as usual, albeit at the new radius too. We always move the excision surface to as large a radius as possible to avoid numerical issues. This means the excision surface and actual apparent horizon are almost the same, differing in radius by the width of one radial cell at most, keeping the boundary consistently inside the apparent horizon whilst minimising numerical difficulties in the hydrodynamics.

\section{\label{sec:metric_solver}Numerical implementation in metric solver}

\subsection{\label{subsec:applying_boundary_conditions}Application of metric boundary conditions}

The differential operator in Equation \eqref{eqn:elliptic_phi} is the flat space Laplacian. Assuming spherical symmetry, this has a kernel consisting of functions of the form $c_{1}r^{-1} + c_{2}$ where $r$ is the radial coordinate and $c_{1}, c_{2}$ are set by the boundary condition at the excision surface (Equation \eqref{eqn:phi_bc}) and the required asymptotic behaviour $\phi \rightarrow 1$ as $r \rightarrow \infty$. An iterative multipole expansion is used to generate a solution, $\phi$, of Equation \eqref{eqn:elliptic_phi} without explicit boundary conditions (see \citep{Muller_Steinmetz:1995, Dimmelmeier_et_al:2005}). Adding the aforementioned element of the kernel to this solution does not change the LHS of Equation \eqref{eqn:elliptic_phi} since, by definition, these terms vanish under the action of the Laplacian. This method produces a solution to the metric equation which satisfies the required boundary conditions. 
Boundary conditions for $\alpha$ and $\beta^{i}$ are enforced in a similar manner, although the differential operator applied to $\beta^{i}$ has slightly different kernel elements: $c_{1}r^{-2} + c_{2}$. The asymptotic values for $\alpha$ and $\beta^{i}$ are $1$ and $0$ respectively at spatial infinity, as mentioned previously.

\subsection{\label{subsec:log_lapse}Reformulated lapse equation}

The excision scheme is switched on only once the apparent horizon has formed as per the condition in Equation \eqref{eqn:AH_condition}. However, we found that Equations~(\ref{eqn:elliptic_phi}--\ref{eqn:elliptic_beta}) tend to become numerically unstable when the minimum value of $\alpha$ approaches
zero so that the code crashes before an apparent horizon is formed.
For small values of $\alpha$, numerical inaccuracies in the metric solver cause the solution for the lapse to become slightly negative at small radii in subsequent iterations. These are purely numerical effects arising from the discretisation due to the spatial resolution not being very high in the core. Locally negative values of $\alpha$ then immediately destabilise the hydrodynamics solver.

To avoid this issue, Equation~\eqref{eqn:elliptic_alpha} is reformulated in log-space \citep[see also][]{Dimmelmeier_et_al:2005},
\begin{equation}
    \begin{split}
    \Delta \log(\alpha \phi) = & 2 \pi \phi^{-2} \bigg[ E^* + 2S^* + \frac{7 \phi^6 K_{ij} K^{ij}}{16 \pi } \bigg] \\
    & - \bigg( \frac{\partial \log(\alpha \phi)}{\partial r} \bigg)^{2}.
\end{split}
\end{equation}
The Poisson equation is now solved for the term $\log{\alpha \phi}$,
The solution for $\alpha \phi = \exp{(\log{\alpha \phi})}$
is then guaranteed to be strictly positive irrespective of the sign of $\log{\alpha \phi}$, thus precluding a negative lapse function. Transforming to log-space induces an additional term of $( \partial/\partial r \log(\alpha \phi) )^{2}$ which must be included in the source term calculation.

The original scheme and the new formulation should coincide during the collapse before black hole formation when there is no risk of a negative lapse function. In a test run using an $85 M\odot$ black-hole forming progenitor,
we found that, from the start of collapse ($t=0$) to $70 \,\textnormal{ms}$ post-bounce and ${\sim} 150\, \textnormal{ms}$ prior to black hole formation, the two methods produce solutions with a cumulative fractional difference of at most $2 \times 10^{-4}$, i.e., two simulations, otherwise identical except for the lapse solution method, have lapse solutions which differ by $<0.02\%$ after $0.61 \, s$ of simulated time. 
It is theoretically conceivable that numerical inaccuracies, for instance: numerically evaluating the new radial derivative the log-transformation incurs, negatively impact the solution. However, there is no compelling indication of this in our results, e.g., local entropy conservation
and ADM mass conservation prior to excision remains on par with the
standard formulation of the lapse equation. Hence, we conclude that the reformulated lapse equation provides a robust and reliable means to
extend simulations to the point of horizon formation.

\section{\label{sec:hydro}Modifications to the hydrodynamics solver}

The details of the hydrodynamic treatment in CoCoNuT have already been discussed by \citet{Dimmelmeier_Font_Mueller:2002} and \citet{Mueller_Janka_Dimmelmeier:2010}. We will limit our description to any changes that are required for excision and black hole formation.

\subsection{\label{subsec:hydro_boundary}Hydrodynamic boundary conditions}

Placing the excision surface inside the apparent horizon guarantees, in theory, that no hydrodynamic artefacts on the boundary can propagate outwards and pollute the rest of the grid. In practice, sensible values must still be enforced to ensure numerical stability. To minimise changes to the code infrastructure, we implement the boundary conditions at the horizons by populating the cells inside the excision region as ghost zones. 

For most variables, e.g. density, pressure, temperature, it is safe to use constant extrapolation into the excised region, i.e., $\rho_{\textnormal{interior}} = \rho_{\textnormal{boundary}}$ in the case of density. The velocity vector requires slightly more finesse, as a poor choice in boundary conditions has been found to produce artefacts in the solution to the relativistic fluid equations. Specifically, we found that a flat velocity profile inside the excised core based on the velocity on the excision surface produces large velocity oscillations just outside the excised region. In the worst case, this can cause the recovery of the primitive variables to crash, as no solution with $v^2<1$ can be found. However, naive linear or higher-order extrapolation of the infall velocity is not viable either since this can result in unphysical superluminal velocities ($v^2>1$) in the ghost zones.

To suppress oscillations in the infall velocity profile near the black hole, we adopt the following strategy: first, $v^{i}$ on the boundary is set to the minimum velocity in the four zones immediately outside the excision surface. This ensures that any spurious oscillations causing small, or even positive, velocities are not copied to the boundary where they may cause numerical issues. We then extrapolate the 
quantity $\phi^{2} W v^{i}$ 
(essentially the four-velocity components measured in the orthonormalised Eulerian frame) into the ghost zones inside the excision region assuming 
$\phi^{2} W v^{i} \propto r^{-1/2}$. This procedure ensure that the extrapolated infall velocity increases monotonically towards smaller radii, while ensuring $v^2<1$ in the ghost cells, and largely eliminates oscillations in the velocity profile outside the excision surface.

\subsection{\label{subsec:internal_energy}Internal energy scheme}

During and after collapse to a black hole, one encounters infall velocities that are close to the speed of light and also highly supersonic. For such kinetically-dominated flow, hydrodynamic schemes based on the total energy equation are prone to stability problems \citep{Mueller:2020} because the recovery of the internal energy density from the total energy density becomes inaccurate.  We found clear symptoms of such problems in the entropy profiles close to the horizon when using the total energy equation. As a result of an inaccurate determination of the internal energy, entropy conservation is violated in regions of very high infall velocity, resulting to a downward drift of entropy during the infall to the horizon, sometimes compounded by isolated glitches in the entropy profiles from small velocity fluctuations in the infall profile. Eventually, this typically lead to equation of state failures some time after black hole formation.

To address these problems, we also evolve the specific internal energy, $\epsilon$, and overwrite the values of $\epsilon$ from the usual recovery of the primitive variables in regions the Mach number of the flow is very high (see below). Internal energy schemes have been widely used in Newtonian hydrodynamics, and \citet{Wilson:2003} have also derived forms for the specific internal energy evolution equation for use in special and general relativistic hydrodynamics simulations. We use a slight variation of Wilson's scheme, which we derive in the following.

We start with the first law of thermodynamics for adiabatic flow, which
gives the change of $\epsilon$ by $P\, \mathrm{dV}$ work,
\begin{equation}
    d\epsilon = -P d\bigg( \frac{1}{\rho} \bigg)
\end{equation}
where $P$ is the pressure, and $\rho$ is the fluid density. This leads
to the covariant form of the internal energy equation,
\begin{equation}
    \label{eqn:pv_work}
    u^{\mu} \nabla_{\mu} \epsilon = -P u^{\mu} \nabla_{\mu} \frac{1}{\rho}.
\end{equation}
By combining Equation~\eqref{eqn:pv_work} with the continuity equation,
\begin{equation}
    \label{eqn:mass_cons}
    \nabla_{\mu} (\rho u^{\mu} ) = u^{\mu} \nabla_{\mu} \rho + \rho \nabla_{\mu} u^{\mu} = 0,
\end{equation}
where $u$ is the 4-velocity, we can bring the RHS of \eqref{eqn:pv_work}
into a more convenient form,
\begin{equation}
    \frac{\partial \epsilon}{\partial t} + \alpha \hat{v}^{i} \frac{\partial \epsilon}{\partial x^{i}} = -\frac{P}{\sqrt{-g} \rho u^{0}} \bigg[ \frac{\partial \sqrt{-g} u^{0}}{\partial t} + \frac{\partial \sqrt{-g} u^{i}}{\partial x^{i}} \bigg].
\end{equation}
Here $g$ is the determinant of the metric and $\hat{v}^{i} = u^{i} / (\alpha u^{0})$. In order to apply the same higher-order finite-volume discretisation
to the LHS as for the continuity, momentum, and energy equation, we multiply by $\sqrt{\gamma} \rho W$ to cast the LHS in flux-conservative form. For numerical stability and computational convenience, it is also convenient to eliminate the spatial derivative term $\partial \sqrt{-g} u^i/\partial x^i$ containing the four-velocity in favour of a derivative of the baryonic mass flux $\sqrt{-g} \rho W \hat{v}^i$, which is available from the solution of the Riemann problem at cell interfaces,
\begin{equation}
\label{eqn:int_energy}
\begin{split}
    \frac{\partial \sqrt{\gamma} \rho W \epsilon}{\partial t} + \frac{\partial \sqrt{\gamma} \rho W \alpha \hat{v}^{i} \epsilon}{\partial x^{i}} = & -\frac{P}{\rho} \frac{\partial \sqrt{-g} \rho W \hat{v}^{i}}{\partial x^{i}} \\
    & - P \frac{\partial \sqrt{\gamma} W}{\partial t} \\
    & - P \rho \sqrt{\gamma} W \hat{v}^{i} \frac{\partial \rho^{-1}}{\partial x^{i}}.
\end{split}
\end{equation}
Here, $\gamma$ is the determinant of the spatial three-metric, $W$ is the Lorentz factor, and $v^i$ is the three-velocity in the
Eulerian frame, which is related to $\hat{v}^{i}$ as
$\hat{v}^{i} = v^{i} - \beta^{i} / \alpha$ with $v^{i}$. 

We only use the value of $\epsilon$ obtained from 
Equation~\eqref{eqn:int_energy} if the square of the Eulerian
three-velocity satisfies $v^{2} > 350 \, c_{\textnormal{s}}^{2}$ where $c_{\textnormal{s}}$ is the local sound speed, i.e., only in
highly supersonic flow.
The factor 350 has no particular physical significance and has been chosen as a compromise that eliminates spurious entropy losses robustly for several hundred milliseconds in our test run, but still restricts the use of the internal energy equation as much as possible (as it violates the strictly conservative nature of the
original hydrodynamics scheme). We do not achieve perfect entropy conservation; the old method of calculating the internal energy from the total energy produces a drop in entropy while the new method tends to cause an upward drift in entropy close to the black hole.
Such an upward drift is more acceptable than the downward drift found for the original total energy equation, as it does not compromise numerical stability and avoids equation of state failures. Since the drift occurs in the highly supersonic region
close to the black hole, it cannot directly affect the infall further upstream. The only minor effect of an upward drift in
entropy close to the horizon is through artificially enhanced
neutrino emission, which will be discussed later in Section~\ref{subsec:neutrino_lum}.

\subsection{Neutrino boundary conditions}
The fast multigroup neutrino transport scheme described by \citet{Mueller_Janka:2015} originally prescribes an inner boundary condition for the neutrino distribution functions at the centre of the simulation grid. During excision, however, this region is not evolved physically and so this boundary condition would become incorrect. Hence, we have moved the boundary condition to the excision surface and minimally altered the underlying physics.

In the original formulation, by a symmetry argument (zero flux at the origin), the value of ingoing and outgoing distribution functions along a ray at the grid centre must be equal. In the excision regime, the constraint we choose is that the outgoing distribution function is zero at the apparent horizon, i.e. the neutrinos are trapped. This introduces a small change to the boundary condition such that the rightmost term of (A10) in \citet{Mueller_Janka:2015} now reads,
\begin{equation}
    f_{+}(r_\mathrm{AH}) = -\frac{\psi(r_\mathrm{AH})}{\chi(r_\mathrm{AH})+1}.
\end{equation}

Additionally, the neutrino transport is not calculated inside the apparent horizon. No other changes are made to the neutrino transport solver.

\section{\label{sec:tests}Code tests}

\begin{figure*}
    \centering
    \includegraphics[width=\textwidth]{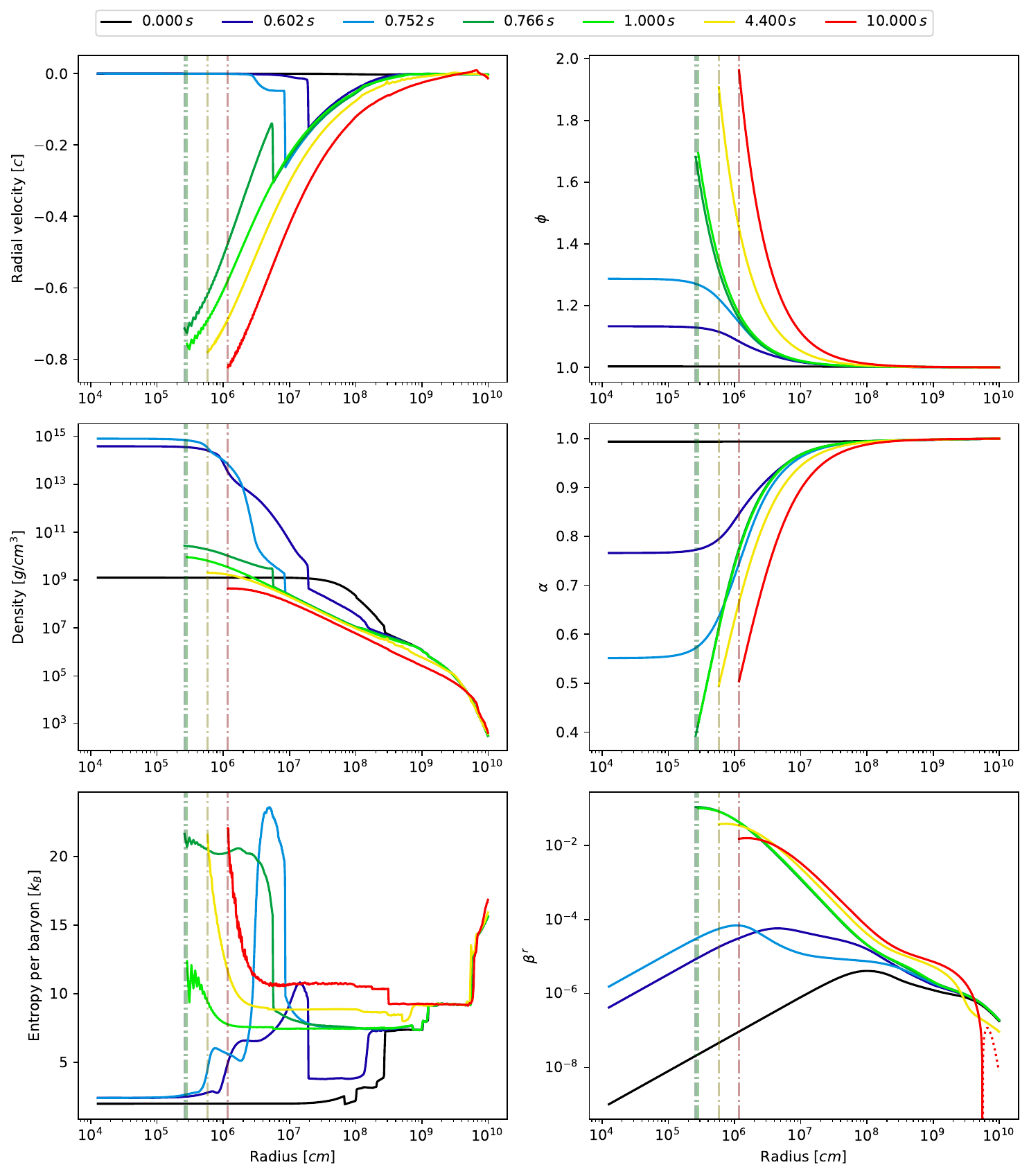}
    \caption{Profiles of the radial velocity (upper left), density (middle left), entropy (lower left), conformal factor ($\phi$; upper right), lapse function ($\alpha$; middle right) and radial component of the shift ($\beta^{r}$; lower right) at various times throughout the 1D simulation. Colours encode the time from the onset of collapse as indicated in the legend at the top of the figure. Only data outside the apparent horizons are shown after black hole formation. The dashed vertical lines show the radius of the excision surface (if it exists) at a time corresponding to the lines' colour.}
    \label{fig:grid_plot}
\end{figure*}

One-dimensional (1D) core-collapse simulations are inherently unable to include important multi-dimensional processes which play a crucial role in the explosion mechanism \citep{Mueller:2020}. However, 1D simulations remain useful, among other purposes, for method verification. We test the
excision scheme in 1D using an $85 {\mathrm M_{\odot}}$ progenitor \citep{Heger_Woosley:2010}. This progenitor is a zero-metallicity (Population III) star which experiences oscillatory instability in oxygen shell burning and pair instability during iron core collapse; see also \citet{Powell_Muller_Heger:2021} for further discussion. For this progenitor, core bounce occurs after $0.54 \, \textnormal{s}$ of simulation time, and the black hole forms after $0.77 \, \textnormal{s}$. In this 1D simulation the shock is not revived and the black hole forms via direct collapse of the core without fallback. The black hole accretes rapidly after it forms, reaching a final (ADM) mass of $15.2 \, M_{\odot}$ at the end of the $10 \, \textnormal{s}$ simulation. The equation of state (EoS) of \citet{Lattimer_Swesty:1991} with a compressibility modulus of $K=220\, \mathrm{MeV}$
is used in the high-density regime. Neutrinos are treated using the fast multigroup transport scheme of \cite{Mueller_Janka:2015}. 
 
Profiles of select hydrodynamic quantities and metric variables are shown in Figure \ref{fig:grid_plot}, depicting important phases of the evolution of the collapsing star.
Before collapse, the shell structure of the progenitor is evident in both the density and entropy profiles from jumps in these quantities at the shell boundaries, which provide convenient markers for the state of the collapse at later times. The velocity profile at this early time shows weak contraction of the entire star as it starts to collapse, with peak infall velocities of $0.03 c$ at a radius of ${\sim} 2 \times 10^3 \, \textnormal{km}$. The spacetime is approximately flat during this period, with $\alpha \approx \phi \approx 1 $ and $\beta^{r} \approx 0$. 

As time progresses, the peak infall velocity increases and moves inwards as the collapsing material falls deeper into the star, correspondingly increasing the density in the central region. The distinctive features of the entropy profiles are advected towards the centre as the layers of the star undergo adiabatic infall. Once the central density reaches some critical value, the collapsing core rebounds, forming a shock clearly visible in the velocity profiles at tens to hundreds of kilometers. 
Shortly thereafter, a characteristic post-bounce entropy profile is established, with a shock-heated PNS
mantle at $\gtrsim 6k_\mathrm{B}/\mathrm{nucleon}$
and higher entropies in the neutrino heating region immediately behind the shock. The lapse and conformal factor now markedly differ from unit. The radial component of the shift vector also increases marginally. 

As material continues to fall onto the PNS, the growing central density drives the lapse lower still while the peak of $\beta^{r}$ increases and moves inwards. Eventually, Equation~\eqref{eqn:AH_condition} signals the formation of an apparent horizon and therefore a black hole. Dashed vertical lines in Figure~\ref{fig:grid_plot} show the position of the excision boundary at each time slice. Data are not shown inside the radii of these lines as the extrapolated solution inside the horizon is not physically meaningful in itself and merely serves to guarantee appropriate boundary conditions at the excision surface. Once the black hole forms, the infall velocity approaches the speed of light outside the excision boundary. At a sufficiently late times, the accretion shock has been advected into the apparent horizon; here the shock discontinuity in the radial velocity, and the post-shock peak in the entropy, are no longer visible. 

After a period of rapid change just prior to black hole formation, the metric slowly adjusts in the subsequent seconds as material accretes onto the black hole. Most significantly, the conformal factor gradually increases with time at any given radius, reflecting the growing mass of the black hole.\footnote{Recall that in vacuum, the mass of static, non-rotating black holes is proportional to the surface area $4\pi \phi^4 r_\mathrm{AH}$ of the horizon.} The value of $\phi$ on the outward-moving horizon surface only increases slowly and stays below the asymptotic value $\phi=2$ at the horizon for the static Schwarzschild spacetime in isotropic coordinates.

As time continues, the lapse function and shift also adjust slightly. One may contrast this post-formation behaviour with the results of \citet{CorderoCarrion_Vasset_Novak_Jaramillo:2014}. In their test simulation, the collapsing body is an unstable neutron star surrounded by vacuum. The physical scale of their simulated collapse is thus much smaller and all material on the grid accretes within tens of milliseconds. In their case, the metric in the long term is stationary, unlike this test where continuous accretion prevents a static solution. 

In the time slice at $10\ \textnormal{s}$, $\beta^{r}$ becomes slightly positive at very large radii. This is shown in Figure~\ref{fig:grid_plot} by the dotted red line representing $-\beta^{r}$ on the logarithmic scale. The negative shift is associated with a positive velocity at the same location and time. The development of a small positive velocity is due to the long duration of the simulation, which becomes comparable to the free-fall timescale of the outermost layers, so that the outer boundary condition for hydrodynamic quantities begins to impact the dynamics of these shells. Overall, the solutions for all metric variables appear stable and show no signs of pathological behaviour. 

In terms of hydrodynamics at late times, the velocity profiles behave as expected with near free-fall infall (Eulerian) velocities that reach a substantial faction of the speed of light at the excision surface. Likewise, the density follows the expected power-law profile ($\rho\propto r ^{-3/2}$) until close to the horizon, where it starts to deviate from the power-law as the infall velocity becomes highly relativistic. The entropy profiles show good conservation of entropy in the infalling matter up to about two times the excision radius, but there is a substantial increase in the immediate vicinity of the horizon surface. This increase is  purely numerical and comes about due to the new prescription for the internal energy described in Section~\ref{subsec:internal_energy} for the strongly subsonic regime.
While this is clearly a sign of limited accuracy in this regime, the increase is preferable to the drop in entropy that would occur if the total energy equation were used at these high infall velocities. Potential consequences of the spurious rise in entropy must be considered carefully (see Section~\ref{subsec:neutrino_lum} below). Numerical noise in the entropy profiles and, to a lesser extent, radial velocity near the excision surface indicate difficulty in the accurate recovery of the primitive variables from the internal energy in the strongly supersonic regime.

The jump in entropy at $r = 3 \times 10^{8} \, \textnormal{cm}$ around $t=10 \, \textnormal{s}$, and dips at previous times at slightly larger radii, are due to nuclear burning. The release of energy increases the entropy of the burned material and produces the sharp feature seen in Figure \ref{fig:grid_plot}.

\begin{figure}
    \centering
    \includegraphics[width=\linewidth]{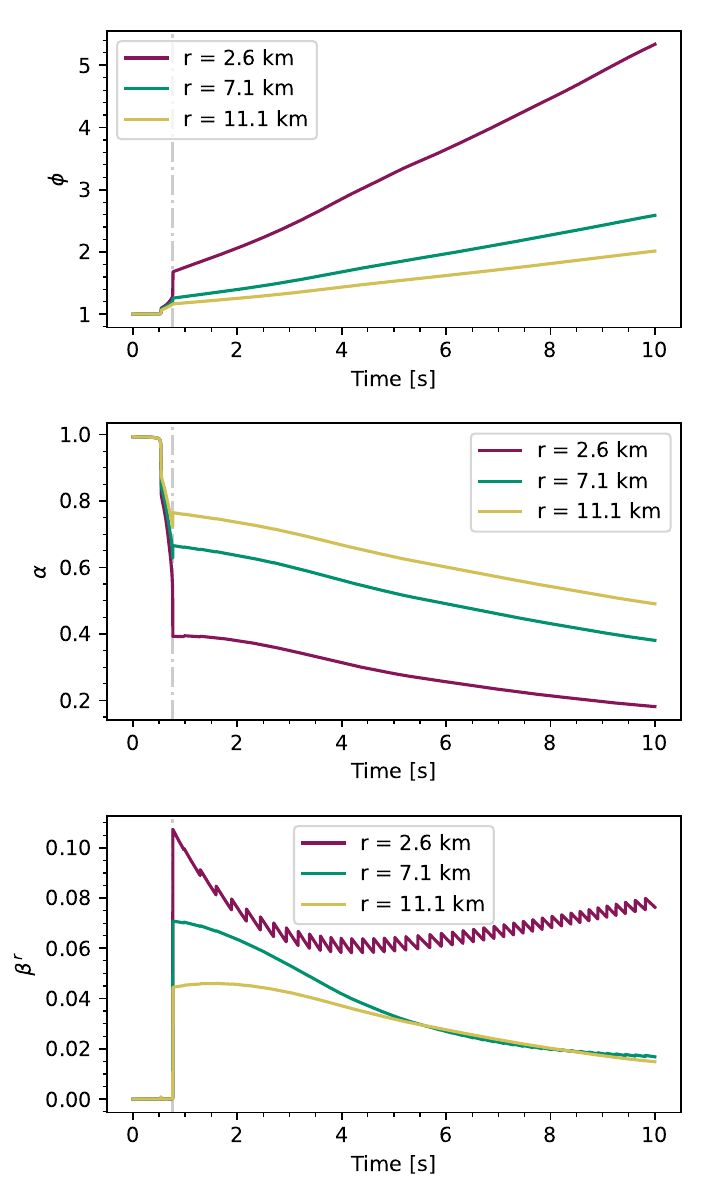}
    \caption{ Time evolution of the metric variables $\phi$ (top), $\alpha$ (middle), and $\beta^{r}$ (bottom) at three different radii. A grey vertical dash-dotted line indicates the time when the black hole forms and the excision scheme is switched on. }
    \label{fig:metric_evolution}
\end{figure}

Figure~\ref{fig:metric_evolution} shows the evolution of metric quantities at three different radii. The radial shift component $\beta^r$ shows discontinuities at times where the excision surface is moved. Such features occur since the boundary condition for the shift vector does not naturally impose continuity in time if the excision radius is moved. It may be possible to remove these jumps in the future using a method which fixes the shift with Equation \eqref{eqn:alpha_bc}, and the lapse with Equation \eqref{eqn:old_alpha_bc}. There are some barely recognisable jumps in the lapse function also, but the method of fixing a ratio of the lapse, $\alpha/(\phi^2 \beta^r)$, each time the excision surface moves minimises any discontinuities. Due to the $1/r$ (for $\phi$ and $\alpha$) and $1/r^{2}$ (for $\beta^{r}$) dependence close to the horizon, these discontinuities appear most prominently at the lowest radius shown (r=$2.6 \textnormal{km}$), which is inside the apparent horizon for almost all of the simulated time after black hole formation. At larger radii, the metric mostly depends smoothly on time with $\phi$ generally increasing, and both $\alpha$ and $\beta^{r}$ tending to decrease. 

\begin{figure*}
    \centering
    \includegraphics[width=\textwidth]{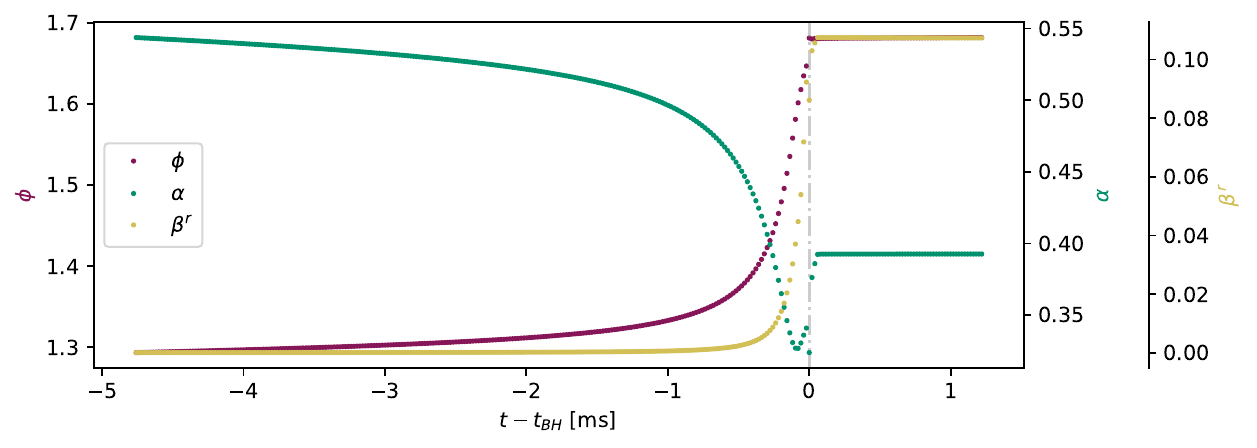}
    \caption{Evolution of the metric very close to the time of BH formation at $r=2.6 \, \textnormal{km}$. Each point represents the output of the simulation at the finest time resolution we computed. That these points form a smooth curve demonstrates the continuity of the solution. The time of black hole formation ($t=0$) is shown by the grey dash-dot line.}
    \label{fig:zoomed_metric}
\end{figure*}

The metric variables initially seem to be discontinuous when the black hole forms. This would be concerning, as continuity of the metric is an important feature. We show in Figure \ref{fig:zoomed_metric} that the sharp jumps in Figure \ref{fig:metric_evolution} at $t=0.77 \, \textnormal{s}$ are primarily due to the very rapid evolution of the metric. Imposing the excision boundary conditions corresponds to noticeable jumps in the metric, however these are small enough to not be a concern. Furthermore, these jumps would appear smaller if each simulation iteration was saved (which is not the case for CoCoNuT), instead tens to hundreds of iterations (where $dt$ is $\mathcal{O} (10^{-7} \, \textnormal{s})$) may pass between outputs, contributing to any apparent discontinuities.

\begin{figure}
    \centering
    \includegraphics[width=\linewidth]{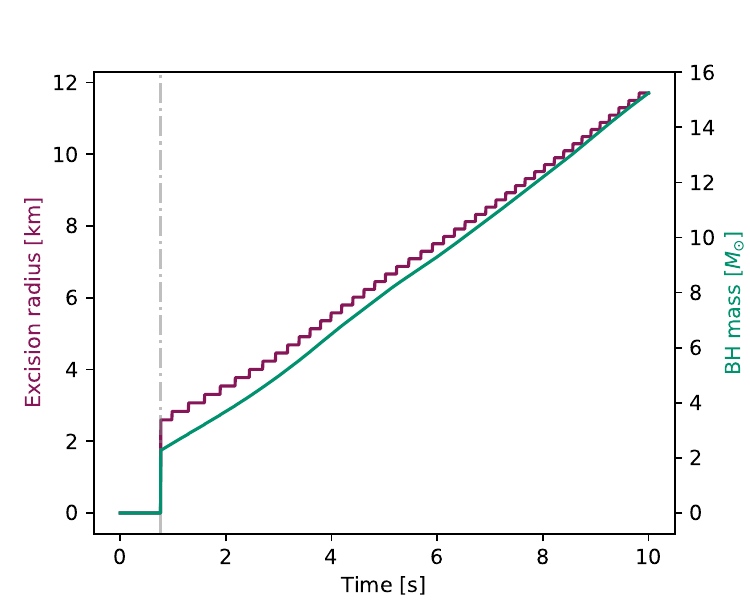}
    \caption{Left axis (purple): Evolution of the excision radius with time. The step-like features are a result of moving the excision surface on a grid with finite spatial resolution. Note that the position of the apparent horizon is not explicitly calculated, but it guaranteed to be within one grid spacing (a quantity which varies approximately logarithmically with radius) of the excision radius. Right axis (green): Formal black hole mass (or more precisely ADM mass enclosed within the excision region) as a function of time.}
    \label{fig:excision_radius}
\end{figure}

While the apparent horizon is initially detected at a radius of $2.6 \, \textnormal{km}$, it gradually moves outwards as material accretes onto the black hole throughout the simulation (see Figure~\ref{fig:excision_radius}). After $10\,  \textnormal{s}$ of simulated time, the excision surface is at a radius of $11.1 \, \textnormal{km}$ with the apparent horizon at a similar but slightly larger radius. Figure~\ref{fig:excision_radius} also shows the mass evolution of the black hole. Here we equate the black hole mass to the ADM mass of the excised region; that is, the ADM mass assuming a vacuum outside the excision region.

\subsection{\label{subsec:adm}ADM mass}

The numerical conservation of the ADM mass is a critical indicator for the accuracy of the evolution of the spacetime metric. In the absence of any energy flux out of the system, the ADM should remain constant in time in the 3+1 formalism \citep{Gourgoulhon:2007} and is an important indicator for the accuracy of the numerical solution of the metric equations.

\begin{figure}
    \centering
    \includegraphics[width=1\linewidth]{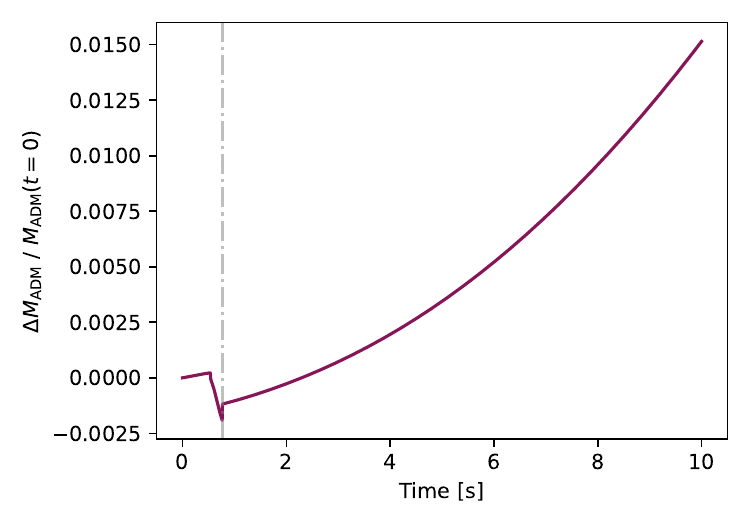}
    \caption{Relative variation of the ADM mass as a function of time with respect to its initial value (i.e., $t=0\, \mathrm{s}$). The time of excision is marked by a vertical dashed grey line. Recall that, by construction, the ADM mass is almost perfectly conserved after excision is switched on except for energy fluxes entering or leaving the grid.}
    \label{fig:adm_mass}
\end{figure}

Figure~\ref{fig:adm_mass} shows the fractional change in the ADM mass over the simulation relative to its initial value. By construction, the ADM mass post-excision changes only slightly with a small positive gradient after excision. Most of this gradient can be attributed to the matter flux at the outer boundary. A much smaller amount is due to the neutrino energy flux, also across the outer boundary (see Equation \eqref{eqn:delta_adm}). From the start of excision to $t=10 \, \textnormal{s}$ the ADM mass changes only in response to these manual energy inputs and outputs and demonstrates the excellent performance of the ADM-conserving scheme. 

\begin{figure}
    \centering
    \includegraphics[width=1\linewidth]{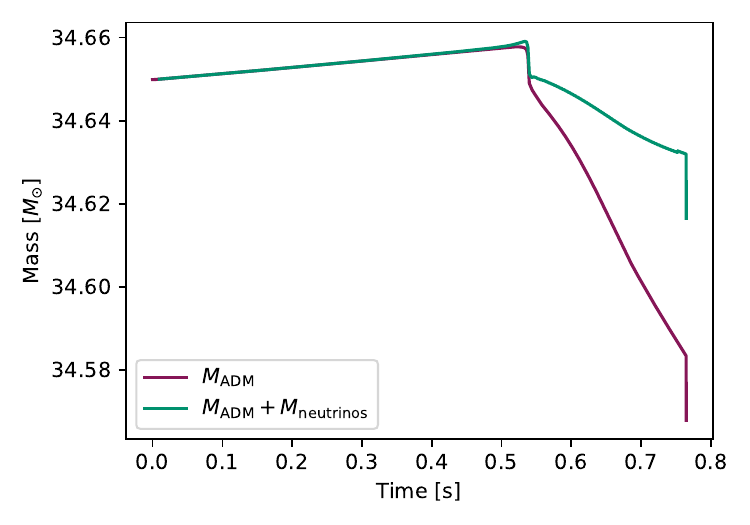}
    \caption{Zoom-in on the evolution of the ADM mass (purple line) before excision, showing relatively small changes as the core collapses. The green line adds to the ADM mass the equivalent time-integrated mass/energy of neutrinos which have propagated across the outer grid boundary. These neutrinos partially explain the decrease in ADM mass before excision.}
    \label{fig:adm_mass_neutrinos}
\end{figure}

Figure~\ref{fig:adm_mass_neutrinos} zooms in on the evolution of the ADM prior to excision. There is a small increase in ADM mass prior to core bounce, followed by a $0.3\%$ drop before the black hole forms. Most of the decrease in ADM mass is due to strong neutrino emission. The green line in Figure~\ref{fig:adm_mass_neutrinos} shows the time-integrated energy of radiated neutrinos added to the ADM mass on the grid. While neutrinos explain most of the ADM mass decrease, there is still a residual drift of numerical origin.

\subsection{\label{subsec:neutrino_lum}Effect of entropy errors on neutrino luminosity}
The increase of entropy just outside the excision surface could potentially alter the dynamics of the collapse. Due to the supersonic nature of the infall, the numerical increase cannot affect the hydrodynamics in the region further upstream directly. However, the dynamics could be altered by the artificial increase in the neutrino energy fluxes that result from an overestimation of the entropy and temperature near the horizon, which would then lead to spurious additional neutrino heating at larger radii. Naturally the prediction of neutrino observables could also be affected.

We investigate the impact of the overestimated entropy on the neutrino signal by flattening the entropy profile that is fed into the neutrino transport solver (via extrapolation from outside the flattened region) once the shock has been advected into the black hole. The entire process for this test works as follows: First, the neutrino transport is solved with no manual change to the entropy, resulting in an update to the neutrino components of the hydrodynamic source terms. This ensures that a valid hydrodynamic update is available for the next timestep, and that the reprocessed neutrino transport solution (computed in a subsequent step) does not impact the dynamics. Next, the entropy is flattened by extrapolating inwards by a constant value from the entropy at a radius where the artificial heating is negligible. The neutrino transport is then solved again, this time without recomputing the source terms but updating and recording the neutrino luminosities. This assumes (correctly, as will be shown later) that neutrinos have a strongly diminished impact on the dynamics of the collapse when artificial heating becomes significant. This method prevents any excess neutrino heating in the gain region from anomalously stronger neutrino emission in the vicinity of the apparent horizon. The hydrodynamic solution is not altered explicitly, i.e., the consequences of the overestimated neutrino emission are considered in isolation, with no feedback into the simulation.

\begin{figure}
    \centering
    \includegraphics[width=\linewidth]{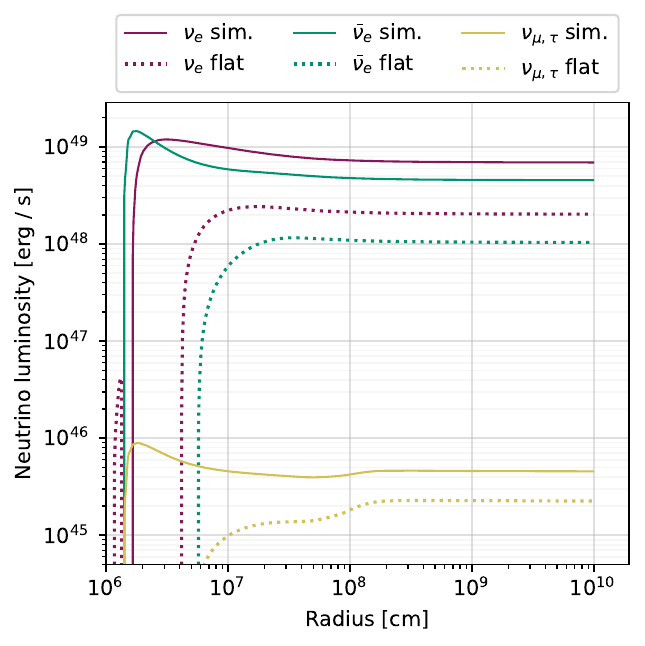}
    \caption{Radial profiles of the neutrino luminosity for the different neutrino species
    $\nu_e$, $\bar{\nu}_e$ and $\nu_{\mu,\tau}$ (denoted by colour) at $t=10 \, \textnormal{s}$. The solid lines show the results of our baseline simulation with a numerical increase in entropy close to the black hole, while the dotted lines show the
    post-processed neutrino luminosity assuming a flat entropy profile near the excised region.}
    \label{fig:neutrino_profiles}
\end{figure}

Figure~\ref{fig:neutrino_profiles} shows radial profiles of the neutrino luminosity at the end of the simulation ($t = 10 \, \textnormal{s}$), calculated from the neutrino energy flux. Profiles at other times exhibit very similar characteristics. At smaller radii, the differences in neutrino luminosity between the original simulation results (solid lines) and the case with a flattened entropy profile (dotted lines) are more significant than at larger radii. Additionally, at small radii, the luminosity becomes negative, indicating that the neutrino flux is inward-pointing. Negative luminosities occur further from the core for the flattened entropy profiles; the difference arises because there is less neutrino production with the flattened entropy model, and so the net neutrino flux is dominated by transmission of inward-directed neutrinos produced at larger radii.
Outside the vicinity of the black hole, the difference between the baseline and flattened-entropy luminosities is smaller and amounts to about a factor of $3.5$ for electron neutrinos. Electron antineutrinos differ a little more, by a factor of $4.5$ between the baseline run and the test case with manually enforced entropy conservation, while the heavy-flavour neutrino luminosity is overestimated by a factor of $2$ at the outer boundary. 

\begin{figure*}
    \centering
    \includegraphics[width=\textwidth]{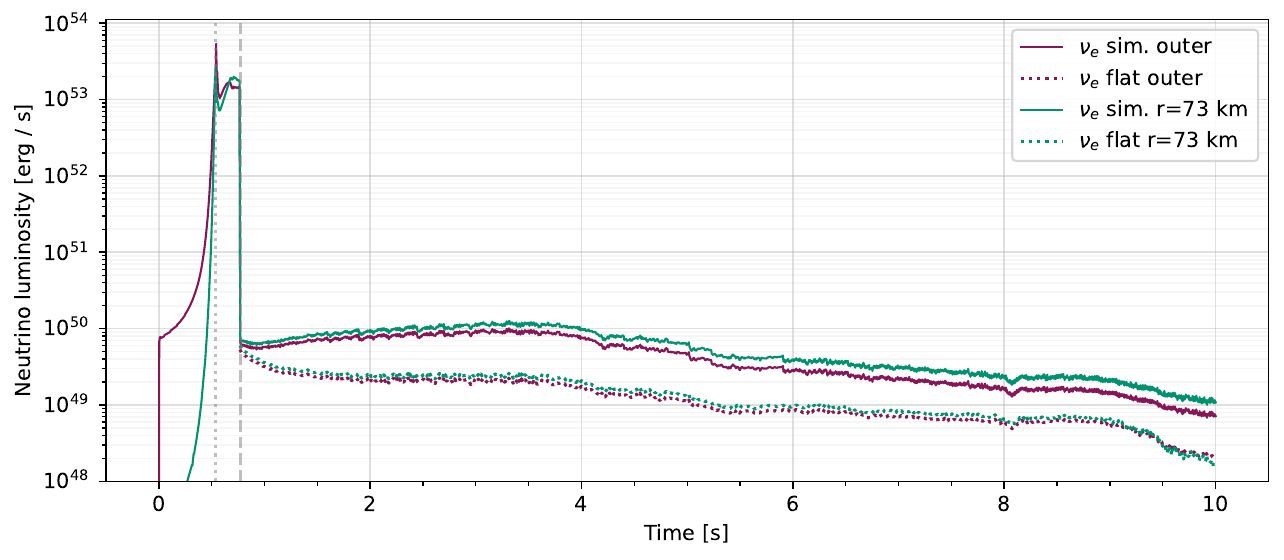}
    \caption{Time evolution of the electron neutrino luminosity at the outer boundary (purple) and near the location of pre-BH shock radius (green) at $73 \, \textnormal{km}$. Solid lines represent simulation results and dotted lines represent post-processed luminosities assuming a flat entropy profile near the horizon. The dotted vertical line indicates the time of core bounce while the dashed vertical line coincides with black hole formation.}
    \label{fig:neutrino_time}
\end{figure*}

The time dependence of the electron neutrino luminosity is shown in Figure~\ref{fig:neutrino_time}. The times of core bounce and black hole formation are marked by a dotted line and a dashed line, respectively. Neutrino luminosities calculated by flattening the entropy profiles outside the excised region once the shock has advected into the black hole (after $t=0.8 \, \textnormal{s}$) are shown by the dotted lines. The purple lines show the luminosity at the outer boundary, a radius of $10^{5} \, \textnormal{km}$, while the green lines show the luminosity at the approximate radius of the shock prior to black hole formation (about $73 \, \textnormal{km}$).

At core bounce, the neutrino luminosity peaks at both radii in the range of a few $10^{53} \, \textnormal{erg} \, \textnormal{s}^{-1}$ as the newly-formed PNS immediately starts cooling. The luminosity remains high until black hole formation, at which point the PNS collapses and the apparent horizon quickly engulfs the neutrinosphere. As a result, the electron neutrino luminosity decreases dramatically by over three orders of magnitude, again at both radii; other neutrino species are similarly affected.\footnote{Note that the drop in neutrino emission manifests itself immediately at the outer boundary due to our use of a stationary neutrino transport scheme, whereas the drop in neutrino emission would in reality take hundreds of milliseconds to manifest itself on the outer grid boundary due to the finite speed of neutrinos.}

The green lines, representing the neutrino luminosity at the approximate shock radius, indicate that, irrespective of the entropy flattening test we performed, neutrino heating in the gain region is abruptly cut off at black hole formation to a level where it becomes dynamically unimportant. Hence there is no concern that the overestimated neutrino luminosities may affect the dynamics of the collapse. The overestimation of the luminosities may still be a concern for the prediction of neutrino observables, but we note that the overestimation only develops shortly \emph{after} the precipitous drop in neutrino emission, i.e., the steep phase of decline at black hole formation, which may still be observable in neutrino detectors, is less affected by the overestimation. In fact, the use of a stationary neutrino transport scheme \citep{Mueller_Janka:2015} is a much more severe approximation when it comes to modelling the abrupt decline in neutrino luminosities. Ideally, one would combine full general relativistic Boltzmann transport and a more accurate hydrodynamic scheme for supersonic infall (e.g., using an entropy evolution equation) to more accurately predict the cut-off of the neutrino signal.

\section{\label{sec:2d}Excision in a 2D Supernova Explosion Model}
The primary purpose of a black hole excision scheme for core-collapse supernovae is to enable simulations of fallback explosions or collapsars, which are inherently multi-dimensional scenarios.
We therefore briefly present a first two-dimensional test simulation with the new excision scheme in this Section.

We consider the same non-rotating $85 M_\odot$ progenitor model as in our 1D test, which has previously been found to undergo shock revival prior to black hole formation in a three-dimensional simulation \citep{Powell_Muller_Heger:2021}. In keeping with the restriction to a spherically symmetric metric, this model is representative of the fallback scenario rather than the collapsar scenario for rapidly rotating progenitors, where deviations of the metric from spherical symmetry and from the conformal flatness condition would become relevant.

To avoid prohibitively small time steps due to the CFL condition, we routinely impose spherical symmetry on the hydrodynamics in multi-dimensions simulation with \textsc{CoCoNuT}
in the PNS core region near the centre of the grid. This spherical region is initially set to encompass the innermost $\mathord{\sim}1 \, \textnormal{km}$. Once the apparent horizon forms, the spherical region is expanded concomitant with the excision surface. Outside this region, we still use multi-dimensional hydrodynamics with the spherically symmetric spacetime.

\begin{figure*}
    \centering
    \begin{subfigure}[b]{0.325\textwidth}
        \centering
        \includegraphics[width=\linewidth]{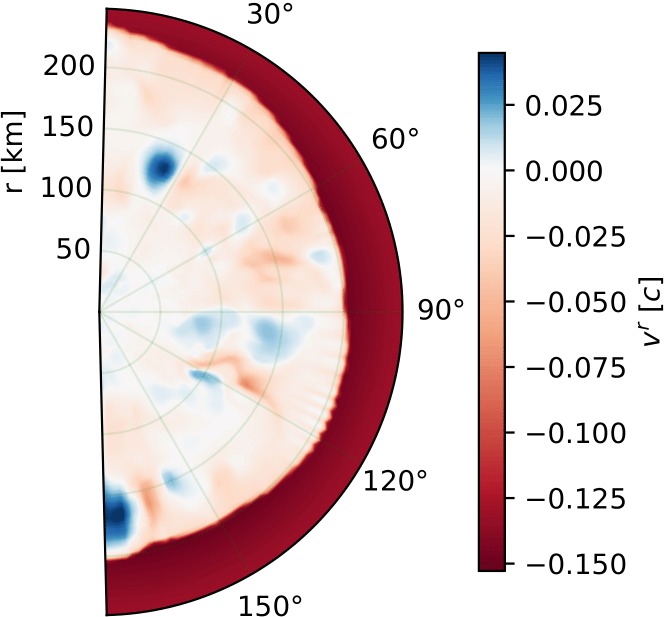}
        \caption{$t=0.06 \, s$}
        \label{subfig:vex_early}
    \end{subfigure}
    \begin{subfigure}[b]{0.325\textwidth}
        \centering
        \includegraphics[width=\linewidth]{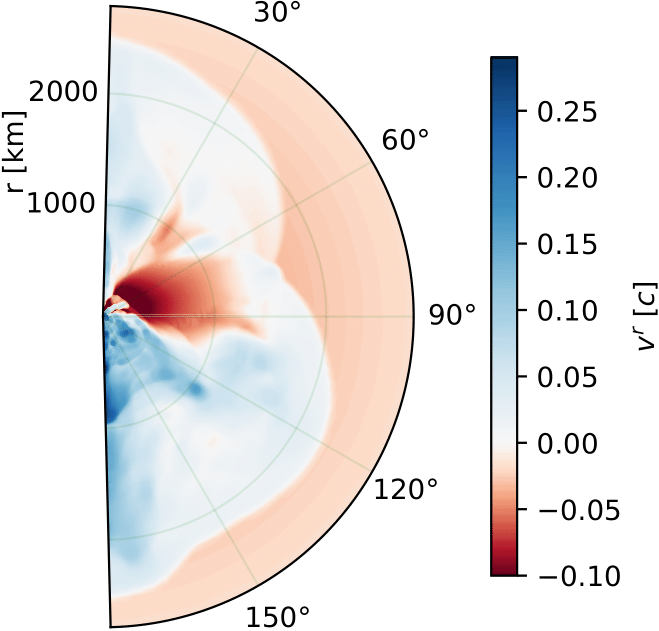}
        \caption{$t = 0.31 \, s$}
        \label{subfig:vex_mid}
    \end{subfigure}
    \begin{subfigure}[b]{0.325\textwidth}
        \centering
        \includegraphics[width=\linewidth]{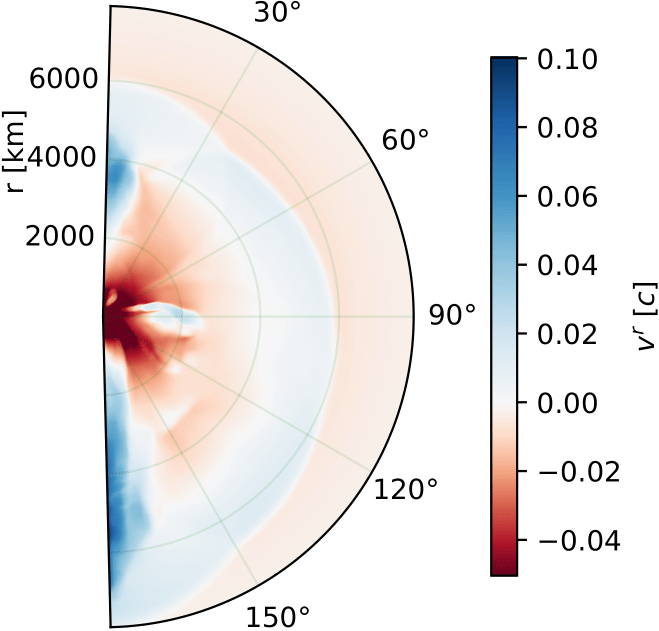}
        \caption{$t = 0.66 \, s$}
        \label{subfig:vex_late}
    \end{subfigure}
    \begin{subfigure}[b]{0.325\textwidth}
        \centering
        \includegraphics[width=\linewidth]{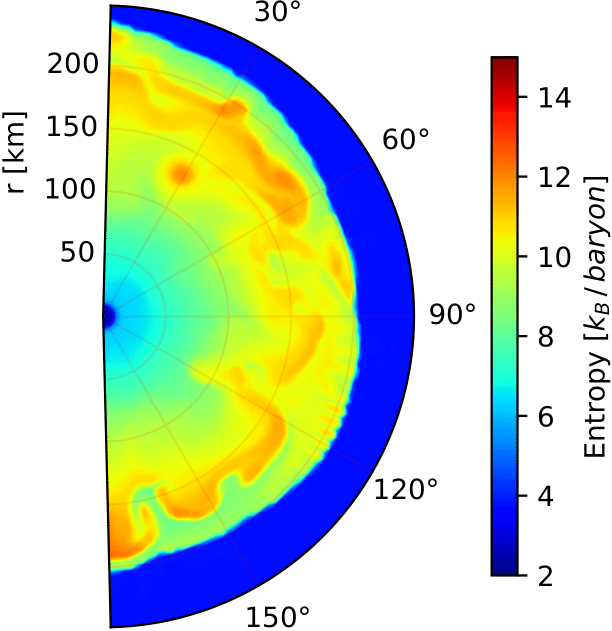}
        \caption{$t=0.06 \, s$}
        \label{subfig:sto_early}
    \end{subfigure}
    \begin{subfigure}[b]{0.325\textwidth}
        \centering
        \includegraphics[width=\linewidth]{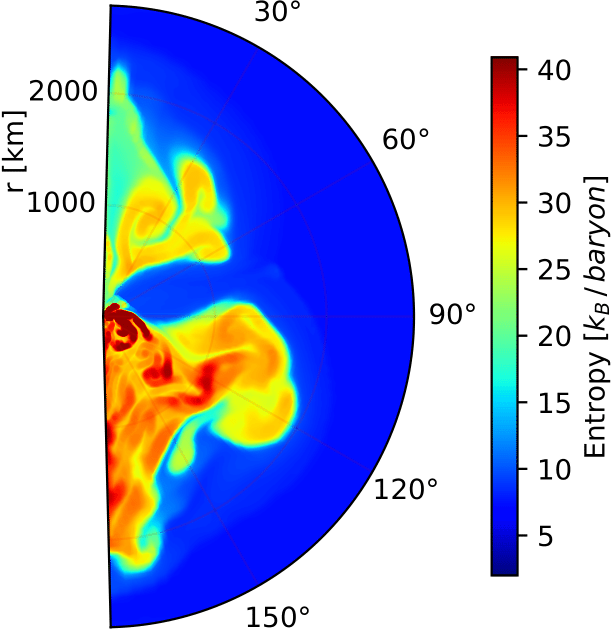}
        \caption{$t = 0.31 \, s$}
        \label{subfig:sto_mid}
    \end{subfigure}
    \begin{subfigure}[b]{0.325\textwidth}
        \centering
        \includegraphics[width=\linewidth]{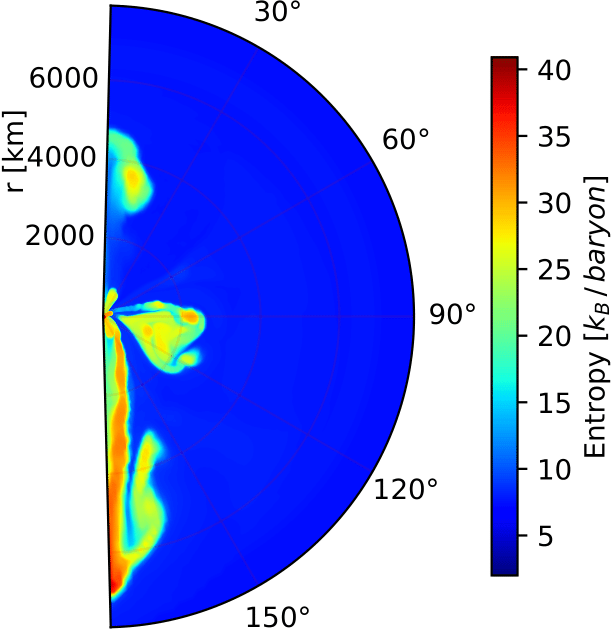}
        \caption{$t = 0.66 \, s$}
        \label{subfig:sto_late}
    \end{subfigure}
    \caption{Snapshots of radial velocity (top row) and entropy (bottom row) during the collapse and fallback explosion of an $85 \, M_{\odot}$ progenitor in 2D. The times shown indicate time post-bounce. Very large negative velocities, $<-0.1\, c$ and $<-0.05 \, c$ for (b) and (c) respectively, are truncated to more clearly show the expanding shock on the colour scale. Note the differing radial axes in each snapshot.}
    \label{fig:2d_results}
\end{figure*}

Figure~\ref{fig:2d_results} shows snapshots from a 2D core-collapse supernovae simulation of the $85 \, M_{\odot}$ progenitor.
The shock forms at $0.54 \, \textnormal{s}$ after the beginning of collapse, the same as in the 1D collapse.
The first snapshot at $t = 0.06 \, \textnormal{s}$ after bounce (Figures~\ref{subfig:vex_early} and
\ref{subfig:sto_early}) shows the stalled shock at a radius of ${\sim}200 \, \textnormal{km}$. At this time, convective overturn has developed with
plumes on medium scales and insignificant global shock deformation.
Whereas the progenitor collapsed to a black hole in 1D
without undergoing shock revival, in 2D, support by convection eventually leads to neutrino-driven shock revival about $0.15\, \mathrm{s}$ after bounce. 
Shock expansion is driven by two prominent neutrino-heated high entropy bubbles (Figures~\ref{subfig:vex_mid} and \ref{subfig:sto_mid}).
The PNS still continues to accrete through a prominent downflow near the equator (albeit at a lower rate than in 1D) while the shock expands, which eventually leads to black hole formation at about $0.31 \, \textnormal{s}$ post-bounce. Because of the reduced accretion rate after shock revival, black hole formation occurs about $80 \, \textnormal{ms}$ later than in 1D.
At this stage, the shock has already expanded to about $2000\, \mathrm{km}$, driven by the hot bubbles along the north and south polar axis. The shock geometry has become strongly bipolar at the time of black hole formation.

Similar to previous simulations that have followed the long-time evolution of fallback explosions after mapping to a Newtonian code at \citep{Chan_Mueller_Heger_Pakmor_Springel:2018,Chan_Mueller:2020} or some time after the collapse of the PNS \citep{Rahman_Janka_Stockinger_Woosley:2021}, the shock and the neutrino-heated bubbles continue to expand after black hole formation, but are continuously decelerated. As a result, much of the material in the hot plumes undergoes fallback onto the black hole. At $0.66\, \mathrm{s}$, three plumes still survive, the biggest of which has expanded to a radius of more than $6000\, \mathrm{km}$ along the South polar axis (Figures~\ref{subfig:vex_late} and \ref{subfig:sto_late}).
$0.35 \, \textnormal{s}$ after formation, the black hole has grown to $2.4 \, M_{\odot}$. In the 1D run, the mass $0.35 \, \textnormal{s}$ after the black hole forms is $2.7 \, M_{\odot}$, i.e., black hole growth in the 2D simulation is delayed somewhat because some of the neutrino-heated material is still expanding and because the infall material from the outer shells is transiently slowed down when it is hit by the shock.

\begin{figure}
    \centering
    \includegraphics[width=\linewidth]{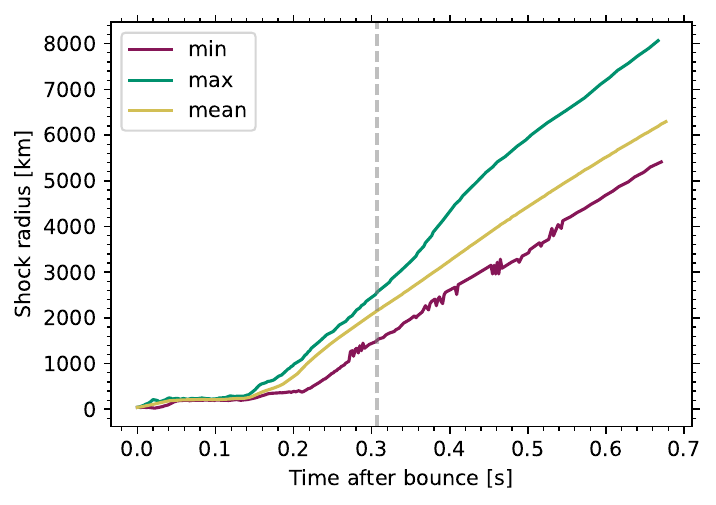}
    \caption{Trajectory of the shock in the 2D fallback supernova simulation of an  $85 \, M_{\odot}$ progenitor as a function of time after bounce. Owing to the asymmetry of the shock we show the minimum, maximum and mean shock radius at each time step. The vertical dashed line marks the time of black hole formation.}
    \label{fig:shock_2d}
\end{figure}

The minimum, maximum and mean of the shock radius over all directions are shown as a summary of the trajectory of the asymmetric shock in Figure \ref{fig:shock_2d}. The shock forms at a radius of tens of kilometers and subsequently expands. It then stalls at approximately $200 \, \textnormal{km}$ for $100 \, \textnormal{ms}$ until sufficient heating produces shock revival. Throughout the explosion phase the shock radius steadily increases, almost linearly with time. The trajectory of the shock shows no immediate response to the formation of the black hole (dashed line in Figure \ref{fig:shock_2d}), however this is expected.

We also examine a diagnostic explosion energy by taking the integral over material with a positive binding energy in line with \citet{Muller_Janka_Marek:2012}. The explosion energy is defined as,

\begin{equation}
    E_{\textnormal{diag}} = \int\limits_{e_{\textnormal{bind}} > 0} e_{\textnormal{bind}} dV,
\end{equation}

where the relativistic binding energy $e_{\textnormal{bind}}$ is given by \citet{Muller_Janka_Marek:2012} and $dV$ is the volume element in curved space (thus incurring an additional $\sqrt{\gamma}$ term).

\begin{figure}
    \centering
    \includegraphics[width=\linewidth]{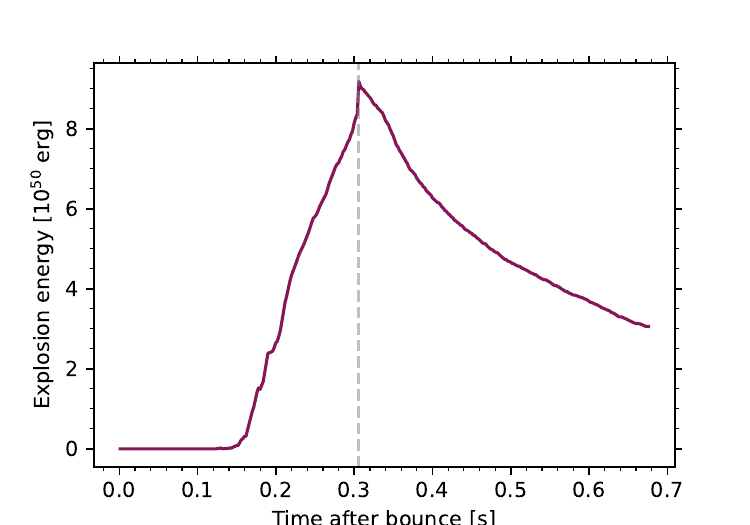}
    \caption{Diagnostic explosion energy for the 2D, fallback supernova model of an $85 \, M_{\odot}$ progenitor. The vertical dashed line marks the time of black hole formation.}
    \label{fig:explosion_2d}
\end{figure}

As shown in Figure \ref{fig:explosion_2d}, material first becomes unbound about $150 \, \textnormal{ms}$ after the shock forms. From here, the explosion energy increases by $>9 \times 10^{50} \, \textnormal{erg}$ in $150 \, \textnormal{ms}$ until black hole formation. There are three main plumes of material with positive binding energy which contribute to the diagnostic explosion energy. As these rise outwards, they exchange energy with their surroundings, resulting in a decrease in their binding energy.

\begin{figure}
    \centering
    \includegraphics[width=\linewidth]{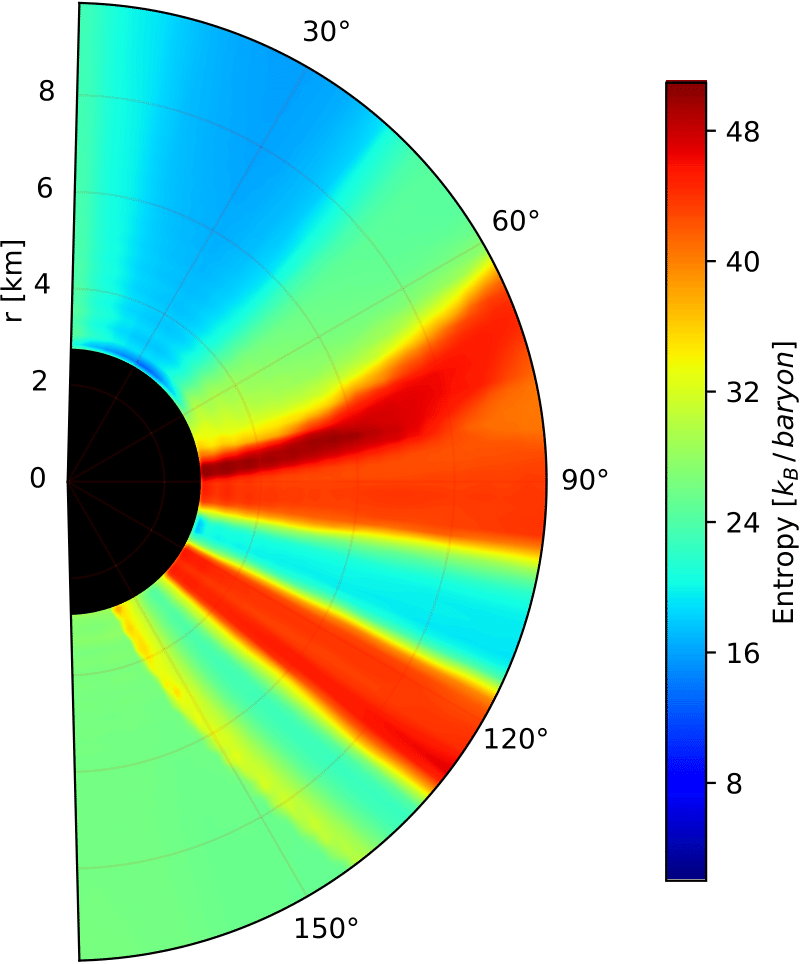}
    \caption{Entropy in the inner $10 \, \textnormal{km}$ of the grid immediately after black hole formation at $t=0.85$. The black area denotes the excised region.}
    \label{fig:sto_zoomed}
\end{figure}

Figure~\ref{fig:sto_zoomed} shows the entropy near the black hole soon after formation. Nearly radial accretion funnels onto the black region, which denotes the excised black hole, display entropies in the range of ${\sim} 20 \texttt{-} 50 \, k_{B} / \textnormal{baryon}$. Like in 1D, 
there is some numerical increase in entropy of the accretion flow, more so in the downflows of moderately warm shocked material (with original post-shock entropies of ${\sim} 20 \, k_{B} / \textnormal{baryon}$) than in hot fallback material from neutrino-heated bubbles, which retains entropies of ${\sim} 20 \, k_{B} / \textnormal{baryon}$ almost down to the apparent horizon. Narrow streaks of higher entropy indicate some localized numerical heating of this material as well.

One can also recognize small boundary artifacts just outside the excision surface about $30 \degree \pm 10 \degree$ from the North pole, where the entropy dips slightly just outside the black hole. However, despite some artifacts, accretion onto the black hole proceeds stably.

\section{\label{sec:conclusion}Conclusion}
We have introduced an excision scheme for core-collapse supernova simulations beyond black hole formation with the neutrino radiation hydrodynamics code \textsc{CoCoNuT-FMT}.
The excision scheme builds on previous work on excision
within a mixed elliptic-hyperbolic formulation of the
Einstein equations by \citet{CorderoCarrion_Vasset_Novak_Jaramillo:2014}, whose excision formalism has been modified slightly to fit specific requirements of long-time core-collapse supernova simulations, such as good ADM mass conservation.
Like the excision scheme of \citet{CorderoCarrion_Vasset_Novak_Jaramillo:2014}, our method is currently limited to a spherically symmetric spacetime metric, but it can be coupled to multi-D hydrodynamics as a first approximation for realistic problems involving black hole formation.
We also discussed modifications to the hydrodynamics solver in the \textsc{CoCoNuT-FMT} code that are required to deal with the strongly relativistic accretion flow onto a black hole. These modifications include the use of the relativistic internal energy equation in the highly supersonic regime as well as carefully chosen boundary condition in the ghost cells inside the excision surface.

We have tested the excision code in spherical symmetry 
and have also performed a 2D simulation of black hole formation in a fallback supernova as a proof of principle.
In 1D, the scheme is able to stably evolve the collapse of a massive star for several seconds after black hole formation. As ADM mass conservation is imposed explicitly via the boundary conditions for the conformal factor, it can accurately follow the growth of the black hole
from $2.3 \, M_{\odot}$ to $14.4 \, M_{\odot}$. However, the highly supersonic accretion flow near the excision surface presents some challenges. While using the internal energy equation instead of the total energy equation avoids stability problems, this comes at the price of an upward drift in entropy in the vicinity of black hole.

Our simulations show limitations in several areas, but the most striking is the unphysical behaviour of the entropy near the excision surface. This issue should be addressed at some point in the future to improve these simulations. As a result, neutrino emission after black hole formation can be overestimated by a factor of several. However, this effect does not strongly come into play after the rapid drop of the neutrino luminosity that is associated with black hole formation, and is dynamically unimportant. The use of an entropy equation for highly supersonic flow may provide a possible remedy in future. Further refinements may come about from a better approximation for the neutrino transport problem or a more accurate equation of state. It may also be possible to relax the spherical symmetry assumption on the spacetime metric in the future. These problems may incur a heavy computational cost to solve and will likely require significant high performance computing resources.

In future, the excision scheme in \textsc{CoCoNuT} will be useful to explore fallback supernovae and the collapsar scenario for long gamma-ray bursts in 2D and 3D simulations. As a proof of principle, we have conducted a 2D simulation of black hole formation in a fallback supernova of a $85 M_\odot$ progenitor. The model has been run for more than $0.3 \, \mathrm{s}$ after the formation of the apparent horizon, and conforms to the qualitative features seen in previous fallback supernova simulations \citep{Chan_Mueller_Heger_Pakmor_Springel:2018,Chan_Mueller:2020,Rahman_Janka_Stockinger_Woosley:2021}. Although the energy of the explosion is slowly drained as the shock scoops up bound material, the shock continues to expand at a steady pace.

As the metric equations are currently solved assuming spherical symmetry, the excision scheme is currently most suitable for fallback explosions of progenitors with slow or modest rotations. Applications to black holes with high spin parameter in the collapsar scenario will require a generalisation of the excision scheme to multi-D within the fully-constrained formalism of \citet{Bonazzola_Gourgoulhon_Grandclement_Novak:2004}. Ideally,
the hyperbolic evolution equations for the  non-conformally flat part of the three-metric and the extrinsic curvature should be fully included. This would also allow for the consistent extraction of gravitational waves around and after black hole formation, when the quadrupole formula is no longer suitable, even with relativistic corrections. In the long term, collapsar simulations should also include fully multi-dimensional Boltzmann neutrino transport \citep{sumiyoshi_et_al:2017,Chan_Mueller:2020} and magnetohydrodynamics.

\begin{acknowledgments}
This research is supported by an Australian Government Research Training Program (RTP) Scholarship. BM acknowledges support by ARC Future Fellowship FT160100035. This work is based on simulations performed within computer time allocations from Astronomy Australia Limited's ASTAC scheme, the National Computational Merit Allocation Scheme (NCMAS), and an Australasian Leadership Computing Grant on the NCI NF supercomputer Gadi. This research was supported by resources provided by the Pawsey Supercomputing Centre, with funding from the Australian Government and the Government of Western Australia. 
This work is also supported by the Spanish Agencia Estatal de Investigaci\'on / Ministerio de Ciencia, Innovaci\'on y Universidades (Grants PGC2018-095984-B-I00 and PID2021-125485NB-C21) and the Generalitat Valenciana (Grant PROMETEO/2019/071). This research was partially supported by the Perimeter Institute for Theoretical Physics through the Simons Emmy Noether program. Research at Perimeter Institute is supported by the Government of Canada through the Department of Innovation, Science and Economic Development and by the Province of Ontario through the Ministry of Research and Innovation.
\end{acknowledgments}

\appendix


\section{\label{app:dADM_dt}Time derivative of ADM surface term}
Recall from Section \ref{subsec:adm} that the ADM mass in spherical symmetry can be expressed as,
\begin{equation}
    \label{eqn:app_adm}
    M_\mathrm{ADM} = -2R^{2}\phi'(R).
\end{equation}

Although not practical for calculating the ADM mass, it can be insightful to express this in terms of the sources in Equation \eqref{eqn:elliptic_phi}, the elliptic equation for the conformal factor. Simply integrating and substituting into the ADM mass equation, one gets
\begin{equation}
    M_\mathrm{ADM} = -2 \int_{r_\mathrm{AH}}^{R} r^2 \Delta \phi dr - 2 r_\mathrm{AH}^{2} \phi'(R),
    \label{eq:adm_app}
\end{equation}

where $r_\mathrm{AH}$ and $R$ are the apparent horizon and outer grid boundary radii respectively. Clearly, Equation~\eqref{eqn:elliptic_phi} can be substituted into the integral to find the contribution to the ADM mass from the volume of the star. The first term of the left hand side is the contribution to the ADM mass from the bulk of the star, from the apparent horizon to the outer boundary (the outer boundary is equivalent to infinity for our purposes). The second term represents the contribution to the ADM mass from material inside the apparent horizon. As the ADM mass should be constant, the change in the first term should be matched by an equal and opposite change in the second. We here consider the time derivative of the second term and show its relation to the source terms of the elliptic equations. We denote the second term $M_{P}$ and simply evaluate,
\begin{equation}
    \frac{\partial M_\mathrm{P}}{\partial t} = -2 r_\mathrm{AH}^{2} \bigg(\frac{\partial}{\partial r} \frac{\partial \phi}{\partial t} \bigg)_{r = r_\mathrm{AH}}
\end{equation}

Fortunately, $\partial_{t} \phi$ is given by Equation \eqref{eqn:phi_bc} and hence we can write,

\begin{align}
    \frac{\partial M_\mathrm{P}}{\partial t} =  -2 r_\mathrm{AH}^{2} \bigg(& \beta^{r} \frac{\partial^{2}\phi}{\partial r^2} + \frac{\phi}{6} \frac{\partial^{2}\beta^{r}}{\partial r^2} + \frac{7}{6} \frac{\partial \phi}{\partial r} \frac{\partial \beta^{r}}{\partial r} \\ \nonumber
    & + \frac{\phi}{3r} \frac{\partial \beta^{r}}{\partial r} + \frac{\beta^{r}}{3r} \frac{\partial \phi}{\partial r} - \frac{\phi \beta^{r}}{3r^{2}} \bigg)_{r=r_\mathrm{AH}}.
\end{align}
Some of these terms can be absorbed in Laplacians, resulting in
\begin{align}
    \label{eqn:dadm_dt_part}
    \frac{\partial M_\mathrm{P}}{\partial t} = -2r_\mathrm{AH}^{2} \bigg( & \beta^{r} \Delta \phi - \frac{5\beta^{r}}{3r} \frac{\partial \phi}{\partial r} \\ \nonumber 
    &+ \frac{\phi}{6} \Delta \beta^{r} + \frac{7}{6} \frac{\partial \phi}{\partial r} \frac{\partial \beta^{r}}{\partial r} \bigg)_{r=r_\mathrm{AH}}.
\end{align}

We note that the first and third term in brackets
can be reexpressed in  terms of the sources in $\Delta \phi$ and $\Delta \beta^{r}$,
\begin{align}
\label{eq:dmdt}
    \frac{\partial M_\mathrm{P}}{\partial t} = 2 r_\mathrm{AH}^{2} \bigg( & \frac{2 \pi \beta^{r} E^{*} }{\phi} + \frac{\beta^{r} \phi^{5} K_{ij}K^{ij}}{8} \\ \nonumber
    &+ \frac{5\beta^{r}}{3r} \frac{\partial \phi}{\partial r} - \frac{2 \pi \alpha} {\phi} (S^{*})^{r} \\ \nonumber
    & - \frac{\phi^{11} K^{rr}}{4} \nabla_{r} \frac{\alpha}{\phi^{6}} - \frac{7}{6} \frac{\partial \phi}{\partial r} \frac{\partial \beta^{r}}{\partial r} \bigg)_{r=r_\mathrm{AH}}.
\end{align}
This makes it manifest that some of the contributions  to the rate of change of $M_\mathrm{P}$ -- the first and fourth term in brackets -- are due to the flux of matter across
the excision surface (including passive advection due to a non-vanishing shift vector).

Equation~\eqref{eq:dmdt} effectively describes the mass gained by the black hole. A numerical evaluation of the individual terms illustrates that achieving the correct long-term evolution of the ADM mass requires a near-exact cancellation of terms in the evolution equations that is difficult to achieve in practice. Equation~(\ref{eq:dmdt}) is dominated by the third and fifth terms. 
Since $\partial \phi/\partial r \sim M/ r^2$, the third
term is of order $\beta^r M_\mathrm{P}/r_\mathrm{AH}$ (by including the $r_\mathrm{AH}^{2}$ outside the brackets), i.e., will drive rapid changes of $M_\mathrm{P}$ on the order of a few light-crossing time scales of the black hole if $\beta^r \gtrsim 0.1$, unless it is almost cancelled by the other terms. Because $K^{rr}\sim \beta^r/r$, the fifth term is of a similar order (as confirmed by numerical evaluation).
To maintain the mass of a black hole of a few solar masses over $\mathord{\sim}1\,\mathrm{s}$ to within $10\%$, these two term need to  cancel each other almost exactly (or be cancelled by other terms)  to an accuracy of about $10^{-6}$. Hence it is unsurprising that we find a significant drift of the ADM mass if we do not explicitly force it to be constant (or change according to the energy flux of neutrinos leaving, and material entering, the computational domain).


\bibliography{bibliography} 

\begin{thebibliography}{54}%
\makeatletter
\providecommand \@ifxundefined [1]{%
 \@ifx{#1\undefined}
}%
\providecommand \@ifnum [1]{%
 \ifnum #1\expandafter \@firstoftwo
 \else \expandafter \@secondoftwo
 \fi
}%
\providecommand \@ifx [1]{%
 \ifx #1\expandafter \@firstoftwo
 \else \expandafter \@secondoftwo
 \fi
}%
\providecommand \natexlab [1]{#1}%
\providecommand \enquote  [1]{``#1''}%
\providecommand \bibnamefont  [1]{#1}%
\providecommand \bibfnamefont [1]{#1}%
\providecommand \citenamefont [1]{#1}%
\providecommand \href@noop [0]{\@secondoftwo}%
\providecommand \href [0]{\begingroup \@sanitize@url \@href}%
\providecommand \@href[1]{\@@startlink{#1}\@@href}%
\providecommand \@@href[1]{\endgroup#1\@@endlink}%
\providecommand \@sanitize@url [0]{\catcode `\\12\catcode `\$12\catcode
  `\&12\catcode `\#12\catcode `\^12\catcode `\_12\catcode `\%12\relax}%
\providecommand \@@startlink[1]{}%
\providecommand \@@endlink[0]{}%
\providecommand \url  [0]{\begingroup\@sanitize@url \@url }%
\providecommand \@url [1]{\endgroup\@href {#1}{\urlprefix }}%
\providecommand \urlprefix  [0]{URL }%
\providecommand \Eprint [0]{\href }%
\providecommand \doibase [0]{https://doi.org/}%
\providecommand \selectlanguage [0]{\@gobble}%
\providecommand \bibinfo  [0]{\@secondoftwo}%
\providecommand \bibfield  [0]{\@secondoftwo}%
\providecommand \translation [1]{[#1]}%
\providecommand \BibitemOpen [0]{}%
\providecommand \bibitemStop [0]{}%
\providecommand \bibitemNoStop [0]{.\EOS\space}%
\providecommand \EOS [0]{\spacefactor3000\relax}%
\providecommand \BibitemShut  [1]{\csname bibitem#1\endcsname}%
\let\auto@bib@innerbib\@empty
\bibitem [{\citenamefont {Baade}\ and\ \citenamefont
  {Zwicky}(1934)}]{Baade_Zwicky:1934}%
  \BibitemOpen
  \bibfield  {author} {\bibinfo {author} {\bibfnamefont {W.}~\bibnamefont
  {Baade}}\ and\ \bibinfo {author} {\bibfnamefont {F.}~\bibnamefont {Zwicky}},\
  }\href {https://doi.org/10.1073/pnas.20.5.254} {\bibfield  {journal}
  {\bibinfo  {journal} {Proceedings of the National Academy of Sciences}\
  }\textbf {\bibinfo {volume} {20}},\ \bibinfo {pages} {254} (\bibinfo {year}
  {1934})}\BibitemShut {NoStop}%
\bibitem [{\citenamefont {{Smartt}}(2009)}]{Smartt:2009}%
  \BibitemOpen
  \bibfield  {author} {\bibinfo {author} {\bibfnamefont {S.~J.}\ \bibnamefont
  {{Smartt}}},\ }\href {https://doi.org/10.1146/annurev-astro-082708-101737}
  {\bibfield  {journal} {\bibinfo  {journal} {\araa}\ }\textbf {\bibinfo
  {volume} {47}},\ \bibinfo {pages} {63} (\bibinfo {year} {2009})}\BibitemShut
  {NoStop}%
\bibitem [{\citenamefont {{Smartt}}(2015)}]{Smartt:2015}%
  \BibitemOpen
  \bibfield  {author} {\bibinfo {author} {\bibfnamefont {S.~J.}\ \bibnamefont
  {{Smartt}}},\ }\href {https://doi.org/10.1017/pasa.2015.17} {\bibfield
  {journal} {\bibinfo  {journal} {\pasa}\ }\textbf {\bibinfo {volume} {32}},\
  \bibinfo {eid} {e016} (\bibinfo {year} {2015})}\BibitemShut {NoStop}%
\bibitem [{\citenamefont {{Adams}}\ \emph {et~al.}(2017)\citenamefont
  {{Adams}}, \citenamefont {{Kochanek}}, \citenamefont {{Gerke}}, \citenamefont
  {{Stanek}},\ and\ \citenamefont
  {{Dai}}}]{Adams_Kochanek_Gerke_Stanek_Dai:2017}%
  \BibitemOpen
  \bibfield  {author} {\bibinfo {author} {\bibfnamefont {S.~M.}\ \bibnamefont
  {{Adams}}}, \bibinfo {author} {\bibfnamefont {C.~S.}\ \bibnamefont
  {{Kochanek}}}, \bibinfo {author} {\bibfnamefont {J.~R.}\ \bibnamefont
  {{Gerke}}}, \bibinfo {author} {\bibfnamefont {K.~Z.}\ \bibnamefont
  {{Stanek}}},\ and\ \bibinfo {author} {\bibfnamefont {X.}~\bibnamefont
  {{Dai}}},\ }\href {https://doi.org/10.1093/mnras/stx816} {\bibfield
  {journal} {\bibinfo  {journal} {\mnras}\ }\textbf {\bibinfo {volume} {468}},\
  \bibinfo {pages} {4968} (\bibinfo {year} {2017})}\BibitemShut {NoStop}%
\bibitem [{\citenamefont {{MacFadyen}}\ and\ \citenamefont
  {{Woosley}}(1999)}]{MacFadyen_Woosley:1999}%
  \BibitemOpen
  \bibfield  {author} {\bibinfo {author} {\bibfnamefont {A.~I.}\ \bibnamefont
  {{MacFadyen}}}\ and\ \bibinfo {author} {\bibfnamefont {S.~E.}\ \bibnamefont
  {{Woosley}}},\ }\href {https://doi.org/10.1086/307790} {\bibfield  {journal}
  {\bibinfo  {journal} {\apj}\ }\textbf {\bibinfo {volume} {524}},\ \bibinfo
  {pages} {262} (\bibinfo {year} {1999})}\BibitemShut {NoStop}%
\bibitem [{\citenamefont {{Woosley}}\ and\ \citenamefont
  {{Bloom}}(2006)}]{Woosley_Bloom:2006}%
  \BibitemOpen
  \bibfield  {author} {\bibinfo {author} {\bibfnamefont {S.~E.}\ \bibnamefont
  {{Woosley}}}\ and\ \bibinfo {author} {\bibfnamefont {J.~S.}\ \bibnamefont
  {{Bloom}}},\ }\href {https://doi.org/10.1146/annurev.astro.43.072103.150558}
  {\bibfield  {journal} {\bibinfo  {journal} {\araa}\ }\textbf {\bibinfo
  {volume} {44}},\ \bibinfo {pages} {507} (\bibinfo {year} {2006})}\BibitemShut
  {NoStop}%
\bibitem [{\citenamefont {{Chan}}\ \emph {et~al.}(2018)\citenamefont {{Chan}},
  \citenamefont {{M{\"u}ller}}, \citenamefont {{Heger}}, \citenamefont
  {{Pakmor}},\ and\ \citenamefont
  {{Springel}}}]{Chan_Mueller_Heger_Pakmor_Springel:2018}%
  \BibitemOpen
  \bibfield  {author} {\bibinfo {author} {\bibfnamefont {C.}~\bibnamefont
  {{Chan}}}, \bibinfo {author} {\bibfnamefont {B.}~\bibnamefont
  {{M{\"u}ller}}}, \bibinfo {author} {\bibfnamefont {A.}~\bibnamefont
  {{Heger}}}, \bibinfo {author} {\bibfnamefont {R.}~\bibnamefont {{Pakmor}}},\
  and\ \bibinfo {author} {\bibfnamefont {V.}~\bibnamefont {{Springel}}},\
  }\href {https://doi.org/10.3847/2041-8213/aaa28c} {\bibfield  {journal}
  {\bibinfo  {journal} {\apjl}\ }\textbf {\bibinfo {volume} {852}},\ \bibinfo
  {eid} {L19} (\bibinfo {year} {2018})}\BibitemShut {NoStop}%
\bibitem [{\citenamefont {{Chan}}\ and\ \citenamefont
  {{M{\"u}ller}}(2020)}]{Chan_Mueller:2020}%
  \BibitemOpen
  \bibfield  {author} {\bibinfo {author} {\bibfnamefont {C.}~\bibnamefont
  {{Chan}}}\ and\ \bibinfo {author} {\bibfnamefont {B.}~\bibnamefont
  {{M{\"u}ller}}},\ }\href {https://doi.org/10.1093/mnras/staa1666} {\bibfield
  {journal} {\bibinfo  {journal} {\mnras}\ }\textbf {\bibinfo {volume} {496}},\
  \bibinfo {pages} {2000} (\bibinfo {year} {2020})}\BibitemShut {NoStop}%
\bibitem [{\citenamefont {{Stockinger}}\ \emph {et~al.}(2020)\citenamefont
  {{Stockinger}}, \citenamefont {{Janka}}, \citenamefont {{Kresse}},
  \citenamefont {{Melson}}, \citenamefont {{Ertl}}, \citenamefont {{Gabler}},
  \citenamefont {{Gessner}}, \citenamefont {{Wongwathanarat}}, \citenamefont
  {{Tolstov}}, \citenamefont {{Leung}}, \citenamefont {{Nomoto}},\ and\
  \citenamefont {{Heger}}}]{stockinger_20}%
  \BibitemOpen
  \bibfield  {author} {\bibinfo {author} {\bibfnamefont {G.}~\bibnamefont
  {{Stockinger}}}, \bibinfo {author} {\bibfnamefont {H.~T.}\ \bibnamefont
  {{Janka}}}, \bibinfo {author} {\bibfnamefont {D.}~\bibnamefont {{Kresse}}},
  \bibinfo {author} {\bibfnamefont {T.}~\bibnamefont {{Melson}}}, \bibinfo
  {author} {\bibfnamefont {T.}~\bibnamefont {{Ertl}}}, \bibinfo {author}
  {\bibfnamefont {M.}~\bibnamefont {{Gabler}}}, \bibinfo {author}
  {\bibfnamefont {A.}~\bibnamefont {{Gessner}}}, \bibinfo {author}
  {\bibfnamefont {A.}~\bibnamefont {{Wongwathanarat}}}, \bibinfo {author}
  {\bibfnamefont {A.}~\bibnamefont {{Tolstov}}}, \bibinfo {author}
  {\bibfnamefont {S.~C.}\ \bibnamefont {{Leung}}}, \bibinfo {author}
  {\bibfnamefont {K.}~\bibnamefont {{Nomoto}}},\ and\ \bibinfo {author}
  {\bibfnamefont {A.}~\bibnamefont {{Heger}}},\ }\href
  {https://doi.org/10.1093/mnras/staa1691} {\bibfield  {journal} {\bibinfo
  {journal} {\mnras}\ }\textbf {\bibinfo {volume} {496}},\ \bibinfo {pages}
  {2039} (\bibinfo {year} {2020})}\BibitemShut {NoStop}%
\bibitem [{\citenamefont {{Rahman}}\ \emph {et~al.}(2021)\citenamefont
  {{Rahman}}, \citenamefont {{Janka}}, \citenamefont {{Stockinger}},\ and\
  \citenamefont {{Woosley}}}]{Rahman_Janka_Stockinger_Woosley:2021}%
  \BibitemOpen
  \bibfield  {author} {\bibinfo {author} {\bibfnamefont {N.}~\bibnamefont
  {{Rahman}}}, \bibinfo {author} {\bibfnamefont {H.-T.}\ \bibnamefont
  {{Janka}}}, \bibinfo {author} {\bibfnamefont {G.}~\bibnamefont
  {{Stockinger}}},\ and\ \bibinfo {author} {\bibfnamefont {S.}~\bibnamefont
  {{Woosley}}},\ }\href@noop {} {\bibfield  {journal} {\bibinfo  {journal}
  {arXiv e-prints}\ ,\ \bibinfo {eid} {arXiv:2112.09707}} (\bibinfo {year}
  {2021})}\BibitemShut {NoStop}%
\bibitem [{\citenamefont {Thornburg}(1993)}]{Thornburg:1993}%
  \BibitemOpen
  \bibfield  {author} {\bibinfo {author} {\bibfnamefont {J.}~\bibnamefont
  {Thornburg}},\ }\emph {\bibinfo {title} {Numerical relativity in black hole
  spacetimes}},\ \href {https://doi.org/http://dx.doi.org/10.14288/1.0085621}
  {Ph.D. thesis},\ \bibinfo  {school} {University of British Columbia}
  (\bibinfo {year} {1993})\BibitemShut {NoStop}%
\bibitem [{\citenamefont {Brandt}\ and\ \citenamefont
  {Br\"ugmann}(1997)}]{Brandt_Brugmann:1997}%
  \BibitemOpen
  \bibfield  {author} {\bibinfo {author} {\bibfnamefont {S.}~\bibnamefont
  {Brandt}}\ and\ \bibinfo {author} {\bibfnamefont {B.}~\bibnamefont
  {Br\"ugmann}},\ }\href {https://doi.org/10.1103/PhysRevLett.78.3606}
  {\bibfield  {journal} {\bibinfo  {journal} {Phys. Rev. Lett.}\ }\textbf
  {\bibinfo {volume} {78}},\ \bibinfo {pages} {3606} (\bibinfo {year}
  {1997})}\BibitemShut {NoStop}%
\bibitem [{\citenamefont {{Artemova}}\ \emph {et~al.}(1996)\citenamefont
  {{Artemova}}, \citenamefont {{Bjoernsson}},\ and\ \citenamefont
  {{Novikov}}}]{Artemova_Bjoernsson_Novikov:1996}%
  \BibitemOpen
  \bibfield  {author} {\bibinfo {author} {\bibfnamefont {I.~V.}\ \bibnamefont
  {{Artemova}}}, \bibinfo {author} {\bibfnamefont {G.}~\bibnamefont
  {{Bjoernsson}}},\ and\ \bibinfo {author} {\bibfnamefont {I.~D.}\ \bibnamefont
  {{Novikov}}},\ }\href {https://doi.org/10.1086/177084} {\bibfield  {journal}
  {\bibinfo  {journal} {\apj}\ }\textbf {\bibinfo {volume} {461}},\ \bibinfo
  {pages} {565} (\bibinfo {year} {1996})}\BibitemShut {NoStop}%
\bibitem [{\citenamefont {{Marek, A.}}\ \emph {et~al.}(2006)\citenamefont
  {{Marek, A.}}, \citenamefont {{Dimmelmeier, H.}}, \citenamefont {{Janka,
  H.-Th.}}, \citenamefont {{M\"uller, E.}},\ and\ \citenamefont {{Buras,
  R.}}}]{Marek_Dimmelmeier_Janka_Muller_Buras:2006}%
  \BibitemOpen
  \bibfield  {author} {\bibinfo {author} {\bibnamefont {{Marek, A.}}}, \bibinfo
  {author} {\bibnamefont {{Dimmelmeier, H.}}}, \bibinfo {author} {\bibnamefont
  {{Janka, H.-Th.}}}, \bibinfo {author} {\bibnamefont {{M\"uller, E.}}},\ and\
  \bibinfo {author} {\bibnamefont {{Buras, R.}}},\ }\href
  {https://doi.org/10.1051/0004-6361:20052840} {\bibfield  {journal} {\bibinfo
  {journal} {A\&A}\ }\textbf {\bibinfo {volume} {445}},\ \bibinfo {pages} {273}
  (\bibinfo {year} {2006})}\BibitemShut {NoStop}%
\bibitem [{\citenamefont {{Kuroda}}\ \emph {et~al.}(2018)\citenamefont
  {{Kuroda}}, \citenamefont {{Kotake}}, \citenamefont {{Takiwaki}},\ and\
  \citenamefont {{Thielemann}}}]{Kuroda_et_al:2018}%
  \BibitemOpen
  \bibfield  {author} {\bibinfo {author} {\bibfnamefont {T.}~\bibnamefont
  {{Kuroda}}}, \bibinfo {author} {\bibfnamefont {K.}~\bibnamefont {{Kotake}}},
  \bibinfo {author} {\bibfnamefont {T.}~\bibnamefont {{Takiwaki}}},\ and\
  \bibinfo {author} {\bibfnamefont {F.-K.}\ \bibnamefont {{Thielemann}}},\
  }\href {https://doi.org/10.1093/mnrasl/sly059} {\bibfield  {journal}
  {\bibinfo  {journal} {\mnras}\ }\textbf {\bibinfo {volume} {477}},\ \bibinfo
  {pages} {L80} (\bibinfo {year} {2018})}\BibitemShut {NoStop}%
\bibitem [{\citenamefont {{Pan}}\ \emph {et~al.}(2018)\citenamefont {{Pan}},
  \citenamefont {{Liebend{\"o}rfer}}, \citenamefont {{Couch}},\ and\
  \citenamefont {{Thielemann}}}]{Pan_Liebendorfer_Couch_Thielemann:2018}%
  \BibitemOpen
  \bibfield  {author} {\bibinfo {author} {\bibfnamefont {K.-C.}\ \bibnamefont
  {{Pan}}}, \bibinfo {author} {\bibfnamefont {M.}~\bibnamefont
  {{Liebend{\"o}rfer}}}, \bibinfo {author} {\bibfnamefont {S.~M.}\ \bibnamefont
  {{Couch}}},\ and\ \bibinfo {author} {\bibfnamefont {F.-K.}\ \bibnamefont
  {{Thielemann}}},\ }\href {https://doi.org/10.3847/1538-4357/aab71d}
  {\bibfield  {journal} {\bibinfo  {journal} {\apj}\ }\textbf {\bibinfo
  {volume} {857}},\ \bibinfo {eid} {13} (\bibinfo {year} {2018})}\BibitemShut
  {NoStop}%
\bibitem [{\citenamefont {{Pan}}\ \emph {et~al.}(2021)\citenamefont {{Pan}},
  \citenamefont {{Liebend{\"o}rfer}}, \citenamefont {{Couch}},\ and\
  \citenamefont {{Thielemann}}}]{Pan_Liebendorfer_Couch_Thielemann:2021}%
  \BibitemOpen
  \bibfield  {author} {\bibinfo {author} {\bibfnamefont {K.-C.}\ \bibnamefont
  {{Pan}}}, \bibinfo {author} {\bibfnamefont {M.}~\bibnamefont
  {{Liebend{\"o}rfer}}}, \bibinfo {author} {\bibfnamefont {S.~M.}\ \bibnamefont
  {{Couch}}},\ and\ \bibinfo {author} {\bibfnamefont {F.-K.}\ \bibnamefont
  {{Thielemann}}},\ }\href {https://doi.org/10.3847/1538-4357/abfb05}
  {\bibfield  {journal} {\bibinfo  {journal} {\apj}\ }\textbf {\bibinfo
  {volume} {914}},\ \bibinfo {eid} {140} (\bibinfo {year} {2021})}\BibitemShut
  {NoStop}%
\bibitem [{\citenamefont {{Isenberg}}(2008)}]{Isenberg:2008}%
  \BibitemOpen
  \bibfield  {author} {\bibinfo {author} {\bibfnamefont {J.~A.}\ \bibnamefont
  {{Isenberg}}},\ }\href {https://doi.org/10.1142/S0218271808011997} {\bibfield
   {journal} {\bibinfo  {journal} {International Journal of Modern Physics D}\
  }\textbf {\bibinfo {volume} {17}},\ \bibinfo {pages} {265} (\bibinfo {year}
  {2008})}\BibitemShut {NoStop}%
\bibitem [{\citenamefont {{Wilson}}\ \emph {et~al.}(1996)\citenamefont
  {{Wilson}}, \citenamefont {{Mathews}},\ and\ \citenamefont
  {{Marronetti}}}]{Wilson_Mathews_Marronetti:1996}%
  \BibitemOpen
  \bibfield  {author} {\bibinfo {author} {\bibfnamefont {J.~R.}\ \bibnamefont
  {{Wilson}}}, \bibinfo {author} {\bibfnamefont {G.~J.}\ \bibnamefont
  {{Mathews}}},\ and\ \bibinfo {author} {\bibfnamefont {P.}~\bibnamefont
  {{Marronetti}}},\ }\href {https://doi.org/10.1103/PhysRevD.54.1317}
  {\bibfield  {journal} {\bibinfo  {journal} {\prd}\ }\textbf {\bibinfo
  {volume} {54}},\ \bibinfo {pages} {1317} (\bibinfo {year}
  {1996})}\BibitemShut {NoStop}%
\bibitem [{\citenamefont {{Cordero-Carri{\'o}n}}\ \emph
  {et~al.}(2009)\citenamefont {{Cordero-Carri{\'o}n}}, \citenamefont
  {{Cerd{\'a}-Dur{\'a}n}}, \citenamefont {{Dimmelmeier}}, \citenamefont
  {{Jaramillo}}, \citenamefont {{Novak}},\ and\ \citenamefont
  {{Gourgoulhon}}}]{CorderoCarrion_et_al:2009}%
  \BibitemOpen
  \bibfield  {author} {\bibinfo {author} {\bibfnamefont {I.}~\bibnamefont
  {{Cordero-Carri{\'o}n}}}, \bibinfo {author} {\bibfnamefont {P.}~\bibnamefont
  {{Cerd{\'a}-Dur{\'a}n}}}, \bibinfo {author} {\bibfnamefont {H.}~\bibnamefont
  {{Dimmelmeier}}}, \bibinfo {author} {\bibfnamefont {J.~L.}\ \bibnamefont
  {{Jaramillo}}}, \bibinfo {author} {\bibfnamefont {J.}~\bibnamefont
  {{Novak}}},\ and\ \bibinfo {author} {\bibfnamefont {E.}~\bibnamefont
  {{Gourgoulhon}}},\ }\href {https://doi.org/10.1103/PhysRevD.79.024017}
  {\bibfield  {journal} {\bibinfo  {journal} {\prd}\ }\textbf {\bibinfo
  {volume} {79}},\ \bibinfo {eid} {024017} (\bibinfo {year}
  {2009})}\BibitemShut {NoStop}%
\bibitem [{\citenamefont {{Fujibayashi}}\ \emph {et~al.}(2021)\citenamefont
  {{Fujibayashi}}, \citenamefont {{Takahashi}}, \citenamefont {{Sekiguchi}},\
  and\ \citenamefont {{Shibata}}}]{Fujibayashi_et_al:2021}%
  \BibitemOpen
  \bibfield  {author} {\bibinfo {author} {\bibfnamefont {S.}~\bibnamefont
  {{Fujibayashi}}}, \bibinfo {author} {\bibfnamefont {K.}~\bibnamefont
  {{Takahashi}}}, \bibinfo {author} {\bibfnamefont {Y.}~\bibnamefont
  {{Sekiguchi}}},\ and\ \bibinfo {author} {\bibfnamefont {M.}~\bibnamefont
  {{Shibata}}},\ }\href@noop {} {\bibfield  {journal} {\bibinfo  {journal}
  {arXiv e-prints}\ ,\ \bibinfo {eid} {arXiv:2102.04467}} (\bibinfo {year}
  {2021})}\BibitemShut {NoStop}%
\bibitem [{\citenamefont {{Komissarov}}\ and\ \citenamefont
  {{Barkov}}(2007)}]{Komissarov_Barkov:2007}%
  \BibitemOpen
  \bibfield  {author} {\bibinfo {author} {\bibfnamefont {S.~S.}\ \bibnamefont
  {{Komissarov}}}\ and\ \bibinfo {author} {\bibfnamefont {M.~V.}\ \bibnamefont
  {{Barkov}}},\ }\href {https://doi.org/10.1111/j.1365-2966.2007.12485.x}
  {\bibfield  {journal} {\bibinfo  {journal} {\mnras}\ }\textbf {\bibinfo
  {volume} {382}},\ \bibinfo {pages} {1029} (\bibinfo {year}
  {2007})}\BibitemShut {NoStop}%
\bibitem [{\citenamefont {{Barkov}}\ and\ \citenamefont
  {{Komissarov}}(2008)}]{Barkov_Komissarov:2008}%
  \BibitemOpen
  \bibfield  {author} {\bibinfo {author} {\bibfnamefont {M.~V.}\ \bibnamefont
  {{Barkov}}}\ and\ \bibinfo {author} {\bibfnamefont {S.~S.}\ \bibnamefont
  {{Komissarov}}},\ }\href {https://doi.org/10.1111/j.1745-3933.2008.00427.x}
  {\bibfield  {journal} {\bibinfo  {journal} {\mnras}\ }\textbf {\bibinfo
  {volume} {385}},\ \bibinfo {pages} {L28} (\bibinfo {year}
  {2008})}\BibitemShut {NoStop}%
\bibitem [{\citenamefont {{Siegel}}\ \emph {et~al.}(2019)\citenamefont
  {{Siegel}}, \citenamefont {{Barnes}},\ and\ \citenamefont
  {{Metzger}}}]{Siegel_Barnes_Metzger:2019}%
  \BibitemOpen
  \bibfield  {author} {\bibinfo {author} {\bibfnamefont {D.~M.}\ \bibnamefont
  {{Siegel}}}, \bibinfo {author} {\bibfnamefont {J.}~\bibnamefont {{Barnes}}},\
  and\ \bibinfo {author} {\bibfnamefont {B.~D.}\ \bibnamefont {{Metzger}}},\
  }\href {https://doi.org/10.1038/s41586-019-1136-0} {\bibfield  {journal}
  {\bibinfo  {journal} {\nat}\ }\textbf {\bibinfo {volume} {569}},\ \bibinfo
  {pages} {241} (\bibinfo {year} {2019})}\BibitemShut {NoStop}%
\bibitem [{\citenamefont {{Pretorius}}(2007)}]{Pretorius:2007}%
  \BibitemOpen
  \bibfield  {author} {\bibinfo {author} {\bibfnamefont {F.}~\bibnamefont
  {{Pretorius}}},\ }\href@noop {} {\bibfield  {journal} {\bibinfo  {journal}
  {arXiv e-prints}\ ,\ \bibinfo {eid} {arXiv:0710.1338}} (\bibinfo {year}
  {2007})}\BibitemShut {NoStop}%
\bibitem [{\citenamefont {Centrella}\ \emph {et~al.}(2010)\citenamefont
  {Centrella}, \citenamefont {Baker}, \citenamefont {Kelly},\ and\
  \citenamefont {van Meter}}]{Centrella_Baker_Kelly_vanMeter:2010}%
  \BibitemOpen
  \bibfield  {author} {\bibinfo {author} {\bibfnamefont {J.}~\bibnamefont
  {Centrella}}, \bibinfo {author} {\bibfnamefont {J.~G.}\ \bibnamefont
  {Baker}}, \bibinfo {author} {\bibfnamefont {B.~J.}\ \bibnamefont {Kelly}},\
  and\ \bibinfo {author} {\bibfnamefont {J.~R.}\ \bibnamefont {van Meter}},\
  }\href {https://doi.org/10.1103/RevModPhys.82.3069} {\bibfield  {journal}
  {\bibinfo  {journal} {Rev. Mod. Phys.}\ }\textbf {\bibinfo {volume} {82}},\
  \bibinfo {pages} {3069} (\bibinfo {year} {2010})}\BibitemShut {NoStop}%
\bibitem [{\citenamefont {Alcubierre}(2008)}]{Alcubierre:2008}%
  \BibitemOpen
  \bibfield  {author} {\bibinfo {author} {\bibfnamefont {M.}~\bibnamefont
  {Alcubierre}},\ }\href@noop {} {\emph {\bibinfo {title} {Introduction to 3+1
  Numerical Relativity}}},\ \bibinfo {series} {International Series of
  Monographs on Physics}, Vol.\ \bibinfo {volume} {140}\ (\bibinfo  {publisher}
  {Oxford University Press},\ \bibinfo {address} {Oxford},\ \bibinfo {year}
  {2008})\BibitemShut {NoStop}%
\bibitem [{\citenamefont {Baumgarte}\ and\ \citenamefont
  {Shapiro}(2010)}]{Baumgarte_Shapiro:2010}%
  \BibitemOpen
  \bibfield  {author} {\bibinfo {author} {\bibfnamefont {T.~W.}\ \bibnamefont
  {Baumgarte}}\ and\ \bibinfo {author} {\bibfnamefont {S.~L.}\ \bibnamefont
  {Shapiro}},\ }\href {https://doi.org/10.1017/CBO9781139193344} {\emph
  {\bibinfo {title} {Numerical Relativity: Solving Einstein's Equations on the
  Computer}}}\ (\bibinfo  {publisher} {Cambridge University Press},\ \bibinfo
  {year} {2010})\BibitemShut {NoStop}%
\bibitem [{\citenamefont {{Dimmelmeier}}\ \emph {et~al.}(2002)\citenamefont
  {{Dimmelmeier}}, \citenamefont {{Font}},\ and\ \citenamefont
  {{M{\"u}ller}}}]{Dimmelmeier_Font_Mueller:2002}%
  \BibitemOpen
  \bibfield  {author} {\bibinfo {author} {\bibfnamefont {H.}~\bibnamefont
  {{Dimmelmeier}}}, \bibinfo {author} {\bibfnamefont {J.~A.}\ \bibnamefont
  {{Font}}},\ and\ \bibinfo {author} {\bibfnamefont {E.}~\bibnamefont
  {{M{\"u}ller}}},\ }\href {https://doi.org/10.1051/0004-6361:20020563}
  {\bibfield  {journal} {\bibinfo  {journal} {\aap}\ }\textbf {\bibinfo
  {volume} {388}},\ \bibinfo {pages} {917} (\bibinfo {year}
  {2002})}\BibitemShut {NoStop}%
\bibitem [{\citenamefont {{M{\"u}ller}}\ \emph {et~al.}(2010)\citenamefont
  {{M{\"u}ller}}, \citenamefont {{Janka}},\ and\ \citenamefont
  {{Dimmelmeier}}}]{Mueller_Janka_Dimmelmeier:2010}%
  \BibitemOpen
  \bibfield  {author} {\bibinfo {author} {\bibfnamefont {B.}~\bibnamefont
  {{M{\"u}ller}}}, \bibinfo {author} {\bibfnamefont {H.-T.}\ \bibnamefont
  {{Janka}}},\ and\ \bibinfo {author} {\bibfnamefont {H.}~\bibnamefont
  {{Dimmelmeier}}},\ }\href {https://doi.org/10.1088/0067-0049/189/1/104}
  {\bibfield  {journal} {\bibinfo  {journal} {\apjs}\ }\textbf {\bibinfo
  {volume} {189}},\ \bibinfo {pages} {104} (\bibinfo {year}
  {2010})}\BibitemShut {NoStop}%
\bibitem [{\citenamefont {{Bonazzola}}\ \emph {et~al.}(2004)\citenamefont
  {{Bonazzola}}, \citenamefont {{Gourgoulhon}}, \citenamefont
  {{Grandcl{\'e}ment}},\ and\ \citenamefont
  {{Novak}}}]{Bonazzola_Gourgoulhon_Grandclement_Novak:2004}%
  \BibitemOpen
  \bibfield  {author} {\bibinfo {author} {\bibfnamefont {S.}~\bibnamefont
  {{Bonazzola}}}, \bibinfo {author} {\bibfnamefont {E.}~\bibnamefont
  {{Gourgoulhon}}}, \bibinfo {author} {\bibfnamefont {P.}~\bibnamefont
  {{Grandcl{\'e}ment}}},\ and\ \bibinfo {author} {\bibfnamefont
  {J.}~\bibnamefont {{Novak}}},\ }\href
  {https://doi.org/10.1103/PhysRevD.70.104007} {\bibfield  {journal} {\bibinfo
  {journal} {\prd}\ }\textbf {\bibinfo {volume} {70}},\ \bibinfo {eid} {104007}
  (\bibinfo {year} {2004})}\BibitemShut {NoStop}%
\bibitem [{\citenamefont {{Jaramillo}}\ \emph {et~al.}(2004)\citenamefont
  {{Jaramillo}}, \citenamefont {{Gourgoulhon}},\ and\ \citenamefont
  {{Marug{\'a}n}}}]{Jaramillo_Gourgoulhon_Marugan:2004}%
  \BibitemOpen
  \bibfield  {author} {\bibinfo {author} {\bibfnamefont {J.~L.}\ \bibnamefont
  {{Jaramillo}}}, \bibinfo {author} {\bibfnamefont {E.}~\bibnamefont
  {{Gourgoulhon}}},\ and\ \bibinfo {author} {\bibfnamefont {G.~A.}\
  \bibnamefont {{Marug{\'a}n}}},\ }\href
  {https://doi.org/10.1103/PhysRevD.70.124036} {\bibfield  {journal} {\bibinfo
  {journal} {\prd}\ }\textbf {\bibinfo {volume} {70}},\ \bibinfo {eid} {124036}
  (\bibinfo {year} {2004})}\BibitemShut {NoStop}%
\bibitem [{\citenamefont {{Jaramillo}}\ \emph {et~al.}(2008)\citenamefont
  {{Jaramillo}}, \citenamefont {{Gourgoulhon}}, \citenamefont
  {{Cordero-Carri{\'o}n}},\ and\ \citenamefont
  {{Ib{\'a}{\~n}ez}}}]{Jaramillo:2008}%
  \BibitemOpen
  \bibfield  {author} {\bibinfo {author} {\bibfnamefont {J.~L.}\ \bibnamefont
  {{Jaramillo}}}, \bibinfo {author} {\bibfnamefont {E.}~\bibnamefont
  {{Gourgoulhon}}}, \bibinfo {author} {\bibfnamefont {I.}~\bibnamefont
  {{Cordero-Carri{\'o}n}}},\ and\ \bibinfo {author} {\bibfnamefont {J.~M.}\
  \bibnamefont {{Ib{\'a}{\~n}ez}}},\ }\href
  {https://doi.org/10.1103/PhysRevD.77.047501} {\bibfield  {journal} {\bibinfo
  {journal} {\prd}\ }\textbf {\bibinfo {volume} {77}},\ \bibinfo {eid} {047501}
  (\bibinfo {year} {2008})}\BibitemShut {NoStop}%
\bibitem [{\citenamefont {{Vasset}}\ \emph {et~al.}(2009)\citenamefont
  {{Vasset}}, \citenamefont {{Novak}},\ and\ \citenamefont
  {{Jaramillo}}}]{Vasset_Novak_Jaramillo:2009}%
  \BibitemOpen
  \bibfield  {author} {\bibinfo {author} {\bibfnamefont {N.}~\bibnamefont
  {{Vasset}}}, \bibinfo {author} {\bibfnamefont {J.}~\bibnamefont {{Novak}}},\
  and\ \bibinfo {author} {\bibfnamefont {J.~L.}\ \bibnamefont {{Jaramillo}}},\
  }\href {https://doi.org/10.1103/PhysRevD.79.124010} {\bibfield  {journal}
  {\bibinfo  {journal} {\prd}\ }\textbf {\bibinfo {volume} {79}},\ \bibinfo
  {eid} {124010} (\bibinfo {year} {2009})}\BibitemShut {NoStop}%
\bibitem [{\citenamefont {{Cordero-Carri{\'o}n}}\ \emph
  {et~al.}(2014)\citenamefont {{Cordero-Carri{\'o}n}}, \citenamefont
  {{Vasset}}, \citenamefont {{Novak}},\ and\ \citenamefont
  {{Jaramillo}}}]{CorderoCarrion_Vasset_Novak_Jaramillo:2014}%
  \BibitemOpen
  \bibfield  {author} {\bibinfo {author} {\bibfnamefont {I.}~\bibnamefont
  {{Cordero-Carri{\'o}n}}}, \bibinfo {author} {\bibfnamefont {N.}~\bibnamefont
  {{Vasset}}}, \bibinfo {author} {\bibfnamefont {J.}~\bibnamefont {{Novak}}},\
  and\ \bibinfo {author} {\bibfnamefont {J.~L.}\ \bibnamefont {{Jaramillo}}},\
  }\href {https://doi.org/10.1103/PhysRevD.90.044062} {\bibfield  {journal}
  {\bibinfo  {journal} {\prd}\ }\textbf {\bibinfo {volume} {90}},\ \bibinfo
  {eid} {044062} (\bibinfo {year} {2014})}\BibitemShut {NoStop}%
\bibitem [{\citenamefont {{Arnowitt}}\ \emph {et~al.}(1962)\citenamefont
  {{Arnowitt}}, \citenamefont {{Deser}},\ and\ \citenamefont
  {{Misner}}}]{ADM:1962}%
  \BibitemOpen
  \bibfield  {author} {\bibinfo {author} {\bibfnamefont {R.}~\bibnamefont
  {{Arnowitt}}}, \bibinfo {author} {\bibfnamefont {S.}~\bibnamefont
  {{Deser}}},\ and\ \bibinfo {author} {\bibfnamefont {C.~W.}\ \bibnamefont
  {{Misner}}},\ }in\ \href@noop {} {\emph {\bibinfo {booktitle} {Gravitation:
  An Introduction to Current Research (Chap. 7).}}},\ \bibinfo {editor} {edited
  by\ \bibinfo {editor} {\bibnamefont {{Louis Witten}}}}\ (\bibinfo
  {publisher} {{John Wiley \& Sons Inc}},\ \bibinfo {year} {1962})\ p.\
  \bibinfo {pages} {227}\BibitemShut {NoStop}%
\bibitem [{\citenamefont {Lichnerowicz}(1944)}]{Lichnerowicz:1944}%
  \BibitemOpen
  \bibfield  {author} {\bibinfo {author} {\bibfnamefont {A.}~\bibnamefont
  {Lichnerowicz}},\ }\href@noop {} {\bibfield  {journal} {\bibinfo  {journal}
  {J. Math. Pures Appl. (9)}\ }\textbf {\bibinfo {volume} {23}},\ \bibinfo
  {pages} {37} (\bibinfo {year} {1944})}\BibitemShut {NoStop}%
\bibitem [{\citenamefont {{Cook}}\ \emph {et~al.}(1996)\citenamefont {{Cook}},
  \citenamefont {{Shapiro}},\ and\ \citenamefont
  {{Teukolsky}}}]{Cook_Shapiro_Teukolsky:1996}%
  \BibitemOpen
  \bibfield  {author} {\bibinfo {author} {\bibfnamefont {G.~B.}\ \bibnamefont
  {{Cook}}}, \bibinfo {author} {\bibfnamefont {S.~L.}\ \bibnamefont
  {{Shapiro}}},\ and\ \bibinfo {author} {\bibfnamefont {S.~A.}\ \bibnamefont
  {{Teukolsky}}},\ }\href {https://doi.org/10.1103/PhysRevD.53.5533} {\bibfield
   {journal} {\bibinfo  {journal} {\prd}\ }\textbf {\bibinfo {volume} {53}},\
  \bibinfo {pages} {5533} (\bibinfo {year} {1996})}\BibitemShut {NoStop}%
\bibitem [{\citenamefont {{Saijo}}(2005)}]{Saijo:2005}%
  \BibitemOpen
  \bibfield  {author} {\bibinfo {author} {\bibfnamefont {M.}~\bibnamefont
  {{Saijo}}},\ }\href {https://doi.org/10.1103/PhysRevD.71.104038} {\bibfield
  {journal} {\bibinfo  {journal} {\prd}\ }\textbf {\bibinfo {volume} {71}},\
  \bibinfo {eid} {104038} (\bibinfo {year} {2005})}\BibitemShut {NoStop}%
\bibitem [{\citenamefont {{Bauswein}}\ \emph {et~al.}(2014)\citenamefont
  {{Bauswein}}, \citenamefont {{Ardevol Pulpillo}}, \citenamefont {{Janka}},\
  and\ \citenamefont {{Goriely}}}]{Bauswein_Pulpillo_Janka_Goriely:2014}%
  \BibitemOpen
  \bibfield  {author} {\bibinfo {author} {\bibfnamefont {A.}~\bibnamefont
  {{Bauswein}}}, \bibinfo {author} {\bibfnamefont {R.}~\bibnamefont {{Ardevol
  Pulpillo}}}, \bibinfo {author} {\bibfnamefont {H.~T.}\ \bibnamefont
  {{Janka}}},\ and\ \bibinfo {author} {\bibfnamefont {S.}~\bibnamefont
  {{Goriely}}},\ }\href {https://doi.org/10.1088/2041-8205/795/1/L9} {\bibfield
   {journal} {\bibinfo  {journal} {\apjl}\ }\textbf {\bibinfo {volume} {795}},\
  \bibinfo {eid} {L9} (\bibinfo {year} {2014})}\BibitemShut {NoStop}%
\bibitem [{\citenamefont {{Dimmelmeier}}\ \emph {et~al.}(2005)\citenamefont
  {{Dimmelmeier}}, \citenamefont {{Novak}}, \citenamefont {{Font}},
  \citenamefont {{Ib{\'a}{\~n}ez}},\ and\ \citenamefont
  {{M{\"u}ller}}}]{Dimmelmeier_et_al:2005}%
  \BibitemOpen
  \bibfield  {author} {\bibinfo {author} {\bibfnamefont {H.}~\bibnamefont
  {{Dimmelmeier}}}, \bibinfo {author} {\bibfnamefont {J.}~\bibnamefont
  {{Novak}}}, \bibinfo {author} {\bibfnamefont {J.~A.}\ \bibnamefont {{Font}}},
  \bibinfo {author} {\bibfnamefont {J.~M.}\ \bibnamefont {{Ib{\'a}{\~n}ez}}},\
  and\ \bibinfo {author} {\bibfnamefont {E.}~\bibnamefont {{M{\"u}ller}}},\
  }\href {https://doi.org/10.1103/PhysRevD.71.064023} {\bibfield  {journal}
  {\bibinfo  {journal} {\prd}\ }\textbf {\bibinfo {volume} {71}},\ \bibinfo
  {eid} {064023} (\bibinfo {year} {2005})}\BibitemShut {NoStop}%
\bibitem [{\citenamefont {{M{\"u}ller}}\ and\ \citenamefont
  {{Steinmetz}}(1995)}]{Muller_Steinmetz:1995}%
  \BibitemOpen
  \bibfield  {author} {\bibinfo {author} {\bibfnamefont {E.}~\bibnamefont
  {{M{\"u}ller}}}\ and\ \bibinfo {author} {\bibfnamefont {M.}~\bibnamefont
  {{Steinmetz}}},\ }\href {https://doi.org/10.1016/0010-4655(94)00185-5}
  {\bibfield  {journal} {\bibinfo  {journal} {Computer Physics Communications}\
  }\textbf {\bibinfo {volume} {89}},\ \bibinfo {pages} {45} (\bibinfo {year}
  {1995})}\BibitemShut {NoStop}%
\bibitem [{\citenamefont {{Grandcl{\'e}ment}}\ \emph
  {et~al.}(2001)\citenamefont {{Grandcl{\'e}ment}}, \citenamefont
  {{Bonazzola}}, \citenamefont {{Gourgoulhon}},\ and\ \citenamefont
  {{Marck}}}]{Grandclement_Bonazzola_Gourgoulhon_Marck:2001}%
  \BibitemOpen
  \bibfield  {author} {\bibinfo {author} {\bibfnamefont {P.}~\bibnamefont
  {{Grandcl{\'e}ment}}}, \bibinfo {author} {\bibfnamefont {S.}~\bibnamefont
  {{Bonazzola}}}, \bibinfo {author} {\bibfnamefont {E.}~\bibnamefont
  {{Gourgoulhon}}},\ and\ \bibinfo {author} {\bibfnamefont {J.~A.}\
  \bibnamefont {{Marck}}},\ }\href {https://doi.org/10.1006/jcph.2001.6734}
  {\bibfield  {journal} {\bibinfo  {journal} {Journal of Computational
  Physics}\ }\textbf {\bibinfo {volume} {170}},\ \bibinfo {pages} {231}
  (\bibinfo {year} {2001})}\BibitemShut {NoStop}%
\bibitem [{\citenamefont {{Cordero-Carri{\'o}n}}\ \emph
  {et~al.}(2012)\citenamefont {{Cordero-Carri{\'o}n}}, \citenamefont
  {{Cerd{\'a}-Dur{\'a}n}},\ and\ \citenamefont
  {{Ib{\'a}{\~n}ez}}}]{CorderoCarrion_CerdaDuran_MariaIbanez:2012}%
  \BibitemOpen
  \bibfield  {author} {\bibinfo {author} {\bibfnamefont {I.}~\bibnamefont
  {{Cordero-Carri{\'o}n}}}, \bibinfo {author} {\bibfnamefont {P.}~\bibnamefont
  {{Cerd{\'a}-Dur{\'a}n}}},\ and\ \bibinfo {author} {\bibfnamefont {J.~M.}\
  \bibnamefont {{Ib{\'a}{\~n}ez}}},\ }\href
  {https://doi.org/10.1103/PhysRevD.85.044023} {\bibfield  {journal} {\bibinfo
  {journal} {\prd}\ }\textbf {\bibinfo {volume} {85}},\ \bibinfo {eid} {044023}
  (\bibinfo {year} {2012})}\BibitemShut {NoStop}%
\bibitem [{\citenamefont {{Gourgoulhon}}(2007)}]{Gourgoulhon:2007}%
  \BibitemOpen
  \bibfield  {author} {\bibinfo {author} {\bibfnamefont {E.}~\bibnamefont
  {{Gourgoulhon}}},\ }\href@noop {} {\bibfield  {journal} {\bibinfo  {journal}
  {arXiv e-prints}\ ,\ \bibinfo {eid} {gr-qc/0703035}} (\bibinfo {year}
  {2007})}\BibitemShut {NoStop}%
\bibitem [{\citenamefont {{Baumgarte}}\ and\ \citenamefont
  {{Shapiro}}(2003)}]{Baumgarte_Shapiro:2003}%
  \BibitemOpen
  \bibfield  {author} {\bibinfo {author} {\bibfnamefont {T.~W.}\ \bibnamefont
  {{Baumgarte}}}\ and\ \bibinfo {author} {\bibfnamefont {S.~L.}\ \bibnamefont
  {{Shapiro}}},\ }\href {https://doi.org/10.1016/S0370-1573(02)00537-9}
  {\bibfield  {journal} {\bibinfo  {journal} {\physrep}\ }\textbf {\bibinfo
  {volume} {376}},\ \bibinfo {pages} {41} (\bibinfo {year} {2003})}\BibitemShut
  {NoStop}%
\bibitem [{\citenamefont {{M{\"u}ller}}(2020)}]{Mueller:2020}%
  \BibitemOpen
  \bibfield  {author} {\bibinfo {author} {\bibfnamefont {B.}~\bibnamefont
  {{M{\"u}ller}}},\ }\href {https://doi.org/10.1007/s41115-020-0008-5}
  {\bibfield  {journal} {\bibinfo  {journal} {Living Reviews in Computational
  Astrophysics}\ }\textbf {\bibinfo {volume} {6}},\ \bibinfo {eid} {3}
  (\bibinfo {year} {2020})}\BibitemShut {NoStop}%
\bibitem [{\citenamefont {{Wilson}}\ and\ \citenamefont
  {{Mathews}}(2003)}]{Wilson:2003}%
  \BibitemOpen
  \bibfield  {author} {\bibinfo {author} {\bibfnamefont {J.~R.}\ \bibnamefont
  {{Wilson}}}\ and\ \bibinfo {author} {\bibfnamefont {G.~J.}\ \bibnamefont
  {{Mathews}}},\ }\href@noop {} {\emph {\bibinfo {title} {{Relativistic
  Numerical Hydrodynamics}}}}\ (\bibinfo  {publisher} {{Cambridge University
  Press}},\ \bibinfo {year} {2003})\BibitemShut {NoStop}%
\bibitem [{\citenamefont {{M{\"u}ller}}\ and\ \citenamefont
  {{Janka}}(2015)}]{Mueller_Janka:2015}%
  \BibitemOpen
  \bibfield  {author} {\bibinfo {author} {\bibfnamefont {B.}~\bibnamefont
  {{M{\"u}ller}}}\ and\ \bibinfo {author} {\bibfnamefont {H.~T.}\ \bibnamefont
  {{Janka}}},\ }\href {https://doi.org/10.1093/mnras/stv101} {\bibfield
  {journal} {\bibinfo  {journal} {\mnras}\ }\textbf {\bibinfo {volume} {448}},\
  \bibinfo {pages} {2141} (\bibinfo {year} {2015})}\BibitemShut {NoStop}%
\bibitem [{\citenamefont {{Heger}}\ and\ \citenamefont
  {{Woosley}}(2010)}]{Heger_Woosley:2010}%
  \BibitemOpen
  \bibfield  {author} {\bibinfo {author} {\bibfnamefont {A.}~\bibnamefont
  {{Heger}}}\ and\ \bibinfo {author} {\bibfnamefont {S.~E.}\ \bibnamefont
  {{Woosley}}},\ }\href {https://doi.org/10.1088/0004-637X/724/1/341}
  {\bibfield  {journal} {\bibinfo  {journal} {\apj}\ }\textbf {\bibinfo
  {volume} {724}},\ \bibinfo {pages} {341} (\bibinfo {year}
  {2010})}\BibitemShut {NoStop}%
\bibitem [{\citenamefont {{Powell}}\ \emph {et~al.}(2021)\citenamefont
  {{Powell}}, \citenamefont {{M{\"u}ller}},\ and\ \citenamefont
  {{Heger}}}]{Powell_Muller_Heger:2021}%
  \BibitemOpen
  \bibfield  {author} {\bibinfo {author} {\bibfnamefont {J.}~\bibnamefont
  {{Powell}}}, \bibinfo {author} {\bibfnamefont {B.}~\bibnamefont
  {{M{\"u}ller}}},\ and\ \bibinfo {author} {\bibfnamefont {A.}~\bibnamefont
  {{Heger}}},\ }\href {https://doi.org/10.1093/mnras/stab614} {\bibfield
  {journal} {\bibinfo  {journal} {\mnras}\ }\textbf {\bibinfo {volume} {503}},\
  \bibinfo {pages} {2108} (\bibinfo {year} {2021})}\BibitemShut {NoStop}%
\bibitem [{\citenamefont {{Lattimer}}\ and\ \citenamefont
  {{Swesty}}(1991)}]{Lattimer_Swesty:1991}%
  \BibitemOpen
  \bibfield  {author} {\bibinfo {author} {\bibfnamefont {J.~M.}\ \bibnamefont
  {{Lattimer}}}\ and\ \bibinfo {author} {\bibfnamefont {D.~F.}\ \bibnamefont
  {{Swesty}}},\ }\href {https://doi.org/10.1016/0375-9474(91)90452-C}
  {\bibfield  {journal} {\bibinfo  {journal} {\nphysa}\ }\textbf {\bibinfo
  {volume} {535}},\ \bibinfo {pages} {331} (\bibinfo {year}
  {1991})}\BibitemShut {NoStop}%
\bibitem [{\citenamefont {{M{\"u}ller}}\ \emph {et~al.}(2012)\citenamefont
  {{M{\"u}ller}}, \citenamefont {{Janka}},\ and\ \citenamefont
  {{Marek}}}]{Muller_Janka_Marek:2012}%
  \BibitemOpen
  \bibfield  {author} {\bibinfo {author} {\bibfnamefont {B.}~\bibnamefont
  {{M{\"u}ller}}}, \bibinfo {author} {\bibfnamefont {H.-T.}\ \bibnamefont
  {{Janka}}},\ and\ \bibinfo {author} {\bibfnamefont {A.}~\bibnamefont
  {{Marek}}},\ }\href {https://doi.org/10.1088/0004-637X/756/1/84} {\bibfield
  {journal} {\bibinfo  {journal} {\apj}\ }\textbf {\bibinfo {volume} {756}},\
  \bibinfo {eid} {84} (\bibinfo {year} {2012})}\BibitemShut {NoStop}%
\bibitem [{\citenamefont {{Sumiyoshi}}\ \emph {et~al.}(2017)\citenamefont
  {{Sumiyoshi}}, \citenamefont {{Nagakura}}, \citenamefont {{Iwakami}},
  \citenamefont {{Furusawa}}, \citenamefont {{Matsufuru}}, \citenamefont
  {{Imakura}},\ and\ \citenamefont {{Yamada}}}]{sumiyoshi_et_al:2017}%
  \BibitemOpen
  \bibfield  {author} {\bibinfo {author} {\bibfnamefont {K.}~\bibnamefont
  {{Sumiyoshi}}}, \bibinfo {author} {\bibfnamefont {H.}~\bibnamefont
  {{Nagakura}}}, \bibinfo {author} {\bibfnamefont {W.}~\bibnamefont
  {{Iwakami}}}, \bibinfo {author} {\bibfnamefont {S.}~\bibnamefont
  {{Furusawa}}}, \bibinfo {author} {\bibfnamefont {H.}~\bibnamefont
  {{Matsufuru}}}, \bibinfo {author} {\bibfnamefont {A.}~\bibnamefont
  {{Imakura}}},\ and\ \bibinfo {author} {\bibfnamefont {S.}~\bibnamefont
  {{Yamada}}},\ }in\ \href {https://doi.org/10.7566/JPSCP.14.010606} {\emph
  {\bibinfo {booktitle} {14th International Symposium on Nuclei in the Cosmos
  (NIC2016)}}},\ \bibinfo {editor} {edited by\ \bibinfo {editor} {\bibfnamefont
  {S.}~\bibnamefont {{Kubono}}}, \bibinfo {editor} {\bibfnamefont
  {T.}~\bibnamefont {{Kajino}}}, \bibinfo {editor} {\bibfnamefont
  {S.}~\bibnamefont {{Nishimura}}}, \bibinfo {editor} {\bibfnamefont
  {T.}~\bibnamefont {{Isobe}}}, \bibinfo {editor} {\bibfnamefont
  {S.}~\bibnamefont {{Nagataki}}}, \bibinfo {editor} {\bibfnamefont
  {T.}~\bibnamefont {{Shima}}},\ and\ \bibinfo {editor} {\bibfnamefont
  {Y.}~\bibnamefont {{Takeda}}}}\ (\bibinfo {year} {2017})\ p.\ \bibinfo
  {pages} {010606}\BibitemShut {NoStop}%
\end{thebibliography}%

\end{document}